# Self-consistent Maxwell-Pauli theory

Sergey A. Rashkovskiy

*Ishlinsky Institute for Problems in Mechanics of the Russian Academy of Sciences, Vernadskogo Ave., 101/1,*

*Moscow, 119526, Russia*

*E-mail: rash@ipmnet.ru, Tel. +7 906 0318854*

**Abstract.** We show that quantum mechanics can be constructed as a classical field theory that correctly describes all basic quantum effects. We construct the self-consistent Maxwell-Pauli theory, from which the correct spontaneous emission spectrum of the hydrogen atom follows. It is shown that many parameters, such as the spin and intrinsic magnetic moment of the electron, which are considered purely quantum properties of the electron and do not have a classical explanation, have a simple and clear physical meaning in the framework of the classical self-consistent Maxwell-Pauli theory.

1.  Introduction

Almost immediately after the discovery of the Schrödinger equation, the idea arose to create a self-consistent theory based on the Maxwell and Schrödinger equations [1]. Subsequently, attempts were repeatedly made to formally combine the Maxwell equations and the wave equations of quantum mechanics (the Schrödinger, Klein-Gordon, Pauli, Dirac equations) into a single self-consistent classical field theory [2–20]. The idea of such a coupling is that the wave functions $\psi$ (scalar, spinor or bispinor) described by the wave equations allows constructing the real parameters that can be interpreted as electric charge density and electric current density continuously distributed in space. Taking this into account, at least from a formal mathematical point of view, we can talk about some electrically charged material (for example, electron) field $\psi$, continuously distributed in space, which is described by the wave equation, just as the classical electromagnetic field is described by Maxwell equations [21]. According to classical electrodynamics, charges and currents continuously distributed in space, create an electromagnetic field, which, in turn, must act on them, changing the field $\psi$, and, hence, its charge density and current density. Thus, the potentials of the electromagnetic field included in the wave equations of quantum mechanics are a superposition of the potentials of the external electromagnetic field (created by external sources) and the potentials of its own electromagnetic field, created by an electrically charged material field $\psi$.

As a result of such a coupling, a self-consistent system of Maxwell-Schrödinger (Maxwell-Klein-Gordon, Maxwell-Pauli, Maxwell-Dirac) equations should arise, which is closed from a mathematical point of view.

The interest in such a coupling of Maxwell equations and the wave equations of quantum mechanics is due to the fact that if it were successfully implemented, it would allow, at least from a formal point of view, considering the resulting equations as a classical theory of interacting continuous fields: an electromagnetic field and a material (electron) field.

The main difficulty of such a coupling is that even in the simplest case of a hydrogen atom, the wave equation (for example, the Schrödinger equation) will include an additional nonlinear term equal to the potential of its own electrostatic field created by the distributed electric charge of the electron field (electron wave):

$$\varphi_0(\mathbf{r}) = -e \int \frac{\psi^*\psi(t,\mathbf{r}')}{R} dV' \qquad (1)$$

where $R = |\mathbf{r} - \mathbf{r}'|$.

In the conventional linear Schrödinger equation, which is the basis of nonrelativistic quantum mechanics and correctly describes the spontaneous emission spectrum of the hydrogen atom, this term is missing. Simple calculations show that the eigenvalues of the nonlinear Schrödinger equation containing the electrostatic potential (1) differ significantly from the eigenvalues of the linear Schrödinger equation. Thus, a simple formal combination of Maxwell equations and the wave equations of quantum mechanics leads to incorrect results in describing the spontaneous emission spectrum of a hydrogen atom.

At the same time, as shown in [22], the intrinsic electromagnetic field of an electron wave plays a significant role in the process of spontaneous emission of an atom: the reverse action of own electromagnetic field of an electron wave in the process of spontaneous emission leads to a rearrangement of the internal structure of an atom, which is traditionally interpreted as "quantum transitions".

An analysis carried out in [21-26] has shown that all basic properties of the hydrogen atom, such as the spontaneous emission, atom-light interaction, photoelectric effect, thermal radiation etc. can be explained within the framework of the self-consistent Maxwell-Schrödinger theory, if we accept that the electron field does not "feel" its own electrostatic field, but at the same time it "feels" the non-stationary (radiative) component of its own electromagnetic field, which occurs during spontaneous radiation.

Thus, it was shown in [22] that the nonlinear Schrödinger equation, which takes into account the own electromagnetic field of the electron wave and correctly describes the rearrangement of the structure of the hydrogen atom during spontaneous emission, should have the form

$$i\hbar \frac{\partial \psi}{\partial t} = \left[ \frac{1}{2m} \left( \frac{\hbar}{i} \nabla + \frac{e}{c} \mathbf{A}_\Sigma \right)^2 - e(\varphi_\Sigma - \varphi_0) \right] \psi \qquad (2)$$

where

$$\varphi_\Sigma = \frac{e}{r} - e \int \frac{|\psi_{t-R/c}|^2}{R} dV' \qquad (3)$$

$$\mathbf{A}_\Sigma = \frac{1}{c} \int \frac{\mathbf{j}_{t-R/c}}{R} dV' \qquad (4)$$

are the scalar and vector potentials of the total (external, i.e. proton and own, i.e. electron wave) electromagnetic field, obtained by formal solution of Maxwell equations, taking into account the distributed charges and currents of the electron wave; **j** is the electric current density of the electron wave, which is defined by the well-known expression of quantum mechanics [27].

In the case when spontaneous emission can be described by the dipole approximation, equation (2), taking into account (1), (3), and (4), can be transformed into the form [22]

$$i\hbar \frac{\partial \psi}{\partial t} = -\frac{1}{2m} \Delta \psi - \frac{e^2}{r} \psi + \frac{2e}{3c^3} \mathbf{r} \ddot{\mathbf{d}} \psi \qquad (5)$$

where

$$\mathbf{d} = -e \int \mathbf{r} |\psi|^2 dV \qquad (6)$$

is the dipole moment of the electron wave in the hydrogen atom.

Equation (5), (6) is a nonlinear Schrödinger equation, and allows describing many quantum effects [21-26], which are traditionally described in terms of QED using the second quantization technique [28]. So, for example, the rearrangement of the structure of the hydrogen atom during spontaneous emission, which in QED is called "spontaneous transition" and is traditionally explained by a zitterbewegung caused by fluctuations of the QED-vacuum, from the point of view of equation (5), (6) turns out to be a completely trivial effect, which has a simple explanation within the framework of classical electrodynamics [22].

Note that, in fact, writing the Schrödinger equation in the form (2) is nothing more than a renormalization widely used in QED to eliminate divergences and infinite energies. Indeed, the subtraction of the intrinsic electrostatic field from the total electromagnetic field leads to the fact that the intrinsic electrostatic energy of charged matter (in this case, an electron wave) is equal to zero. If this is not done, then for a point particle this will lead to an infinite value of the energy of its interaction with its own electrostatic field, while for the electron wave considered here, to incorrect eigenfrequencies of the hydrogen atom and, as a consequence, to incorrect spontaneous emission spectrum that do not agree with experiment.

Despite the success of equation (2) (or, what is the same, equation (5)) in explaining a number of basic "quantum" regularities [21-26], from the point of view of classical electrodynamics, it remains inexplicable why an electrically charged electron field does not "feel" its own electrostatic field, but at the same time it "feels" the electrostatic field of other charges and the non-stationary (radiative) component of its own electromagnetic field, which occurs during spontaneous emission.

The purpose of this work is to eliminate this "paradox" and analyze the joint system of equations describing the interacting electromagnetic and electron fields in the framework of classical field theory without their quantization.

Note that equations (2) and (5) do not take into account such property of matter as spin.

As shown in [21], from the point of view of classical field theory, an electron field, in addition to the electric charge and electric current continuously distributed in space, also has its own angular momentum (spin) continuously distributed in space and its own magnetic moment associated with spin, which are not reduced to the mechanical movement of charges, while the intrinsic gyromagnetic ratio turns out to be 2 times greater than the gyromagnetic ratio associated with the movement (flow) of electric charges.

In the general case, the electron field is described by the Dirac equation, which for it is a field equation, similar to Maxwell equations which are the field equations for a classical electromagnetic field. However, in this work, we restrict ourselves to the long-wavelength approximation of the Dirac equation, i.e., the Pauli equation, and consider the coupled Maxwell-Pauli equations, which do not have the above disadvantage and allow describing the main "quantum" regularities in the framework of classical field theory without fields quantization.

## 2. Maxwell-Pauli field

The simplest formal generalization of the nonlinear Schrödinger equation (2) taking into account the spin (intrinsic magnetic moment) of the electron field has the form

$$i\hbar \frac{\partial \Psi}{\partial t} = \left[\frac{1}{2m_e}\left(\frac{\hbar}{i}\nabla + \frac{e}{c}\mathbf{A}_\Sigma\right)^2 - e(\varphi_\Sigma - \varphi_0) + \frac{e\hbar}{2m_e c}\boldsymbol{\sigma}\mathbf{H}_\Sigma\right]\Psi \qquad (7)$$

where $\Psi = \begin{pmatrix}\psi_1\\ \psi_2\end{pmatrix}$, $\Psi^* = (\psi_1^* \quad \psi_2^*)$,

$$\varphi_0(\mathbf{r}) = \int \frac{\rho_t}{R} dV' \qquad (8)$$

is the potential of the own electrostatic field of the electron wave, while the scalar and vector potentials of the total electromagnetic field are determined by the known solutions of the Maxwell equations [29]

$$\varphi_\Sigma = \varphi + \int \frac{\rho_{t-R/c}}{R} dV' \qquad (9)$$

$$\mathbf{A}_\Sigma = \mathbf{A} + \frac{1}{c}\int \frac{\mathbf{j}_{t-R/c}}{R} dV' \qquad (10)$$

$$\mathbf{H}_\Sigma = \text{rot}\mathbf{A}_\Sigma \qquad (11)$$

$$\rho = -e\Psi^*\Psi = -e(|\psi_1|^2 + |\psi_2|^2) \tag{12}$$

is the electric charge density;

$$\mathbf{j} = \frac{e\hbar}{2m_e i}[(\nabla\Psi^*)\Psi - \Psi^*\nabla\Psi] - \frac{e^2}{m_e c}\mathbf{A}_\Sigma\Psi^*\Psi - \frac{e\hbar}{2m_e}\text{rot}(\Psi^*\boldsymbol{\sigma}\Psi) \tag{13}$$

is the electric current density of the electron field in the Pauli approximation [28]; $\varphi$ and $\mathbf{A}$ are the scalar and vector potentials of the external electromagnetic field (i.e., the field created by the nucleus of an atom and sources external to the atom); the index $t - R/c$ of the charge density and current density indicates that they are taken at the time $t' = t - R/c$.

Note that the field $\varphi_0(\mathbf{r})$ satisfies the equation

$$\Delta\varphi_0 = -4\pi\rho \tag{14}$$

An obvious disadvantage of equation (7), as well as equation (2), is that the electrostatic potential $\varphi_0$ (8) includes the electric charge density of only the considered electron wave (for which equations (2) and (7) are written), but it does not include the density of external electric charges (for example, the nucleus of the same atom, as well as protons and electron waves of other atoms and ions). This leads to a new paradox: how does an electron wave "distinguish" its own electrostatic field from the electrostatic field of other charges?

Let us show that, along with Eq. (7), an informal generalization of Eq. (2) for the electron field in the Pauli approximation is possible.

Consider the system of equations:

$$i\hbar\frac{\partial\Psi}{\partial t} = \left[\frac{1}{2m_e}\left(\frac{\hbar}{i}\nabla + \frac{e}{c}\mathbf{A}_\Sigma\right)^2 - e\varphi_\Sigma - e\boldsymbol{\sigma}\mathbf{G}_\Sigma + \frac{e\hbar}{2m_e c}\boldsymbol{\sigma}\mathbf{H}_\Sigma\right]\Psi \tag{15}$$

$$\frac{1}{c^2}\frac{\partial^2\mathbf{G}_\Sigma}{\partial t^2} - \Delta\mathbf{G}_\Sigma = 4\pi e(\Psi^*\boldsymbol{\sigma}\Psi) \tag{16}$$

$$\mathbf{H}_\Sigma = \mathbf{H} + \mathbf{H}_e, \mathbf{E}_\Sigma = \mathbf{E} + \mathbf{E}_e, \mathbf{G}_\Sigma = \mathbf{G} + \mathbf{G}_e \tag{17}$$

$$\varphi_\Sigma = \varphi + \varphi_e, \mathbf{A}_\Sigma = \mathbf{A} + \mathbf{A}_e \tag{18}$$

$$\text{rot}\mathbf{H}_\Sigma = \frac{1}{c}\frac{\partial\mathbf{E}_\Sigma}{\partial t} + \frac{4\pi}{c}\mathbf{j} \tag{19}$$

$$\text{div}\mathbf{E}_\Sigma = 4\pi\rho \tag{20}$$

where

$$\mathbf{H}_\Sigma = \text{rot}\mathbf{A}_\Sigma, \ \mathbf{E}_\Sigma = -\frac{1}{c}\frac{\partial\mathbf{A}_\Sigma}{\partial t} - \nabla\varphi_\Sigma \tag{21}$$

$$\rho = -e\Psi^*\Psi \tag{22}$$

$$\mathbf{j} = \frac{e\hbar}{2m_e i}[(\nabla\Psi^*)\Psi - \Psi^*\nabla\Psi] - \frac{e^2}{m_e c}\mathbf{A}_\Sigma\Psi^*\Psi - \frac{e\hbar}{2m_e}\text{rot}(\Psi^*\boldsymbol{\sigma}\Psi) \tag{23}$$

the index "$e$" refers to the own fields of the electron wave described by the spinor $\Psi$; parameters without an index refer to external fields created by external (with respect to the field $\Psi$) charges, currents, and spins. Here it is assumed that in the considered region of space there are no electric charges, currents and spins that create external fields.

Further, we will use synonyms: Pauli field, electron field, electron wave, denoting the same physical object: a classical electrically charged field described by equation (15).

The system of equations (15)-(23) is closed and self-consistent. It differs from a simple formal union of the Maxwell and Pauli equations in that the Pauli equation (15) includes an additional term $-e\boldsymbol{\sigma}\mathbf{G}_\Sigma$ containing a new real vector field $\mathbf{G}_\Sigma$ satisfying equation (16). As we will see below, this fundamentally changes the solutions of the combined system of Maxwell-Pauli equations, and allows correctly describing the experimentally observed effects. Below, we will use synonyms: **G**-field and spin field (not to be confused with the spinor field!), denoting the same physical object: the classical vector field created by the spin (spin magnetic moment) of the Pauli field and described by equation (16).

Note that the introduction of the term $-e\boldsymbol{\sigma}\mathbf{G}_\Sigma$ into Eq. (15) can be considered as a renormalization of the theory; however, in contrast to the formal (artificial) renormalization which is introduced in QED, this renormalization has a simple and natural meaning, as the effect of some vector field $\mathbf{G}_\Sigma$ satisfying equation (16), which was not previously taken into account in the theory.

It is easy to see that the system of equations (15)-(23) has gauge invariance:

$$\mathbf{A}_\Sigma \to \mathbf{A}_\Sigma + \nabla f, \ \varphi_\Sigma \to \varphi_\Sigma - \frac{1}{c}\frac{\partial f}{\partial t}, \ \Psi \to \Psi \exp\left(-\frac{ie}{\hbar c}f\right) \qquad (24)$$

where $f$ is an arbitrary function; in this case, the strengths of the electric and magnetic fields, the field $\mathbf{G}_\Sigma$, the density of the electric charge, the current density and other physical characteristics of the electron field do not change.

It follows from equations (19) and (20) that $\rho$ and **j** defined by relations (22) and (23) really play the role of the electric charge density and electric current density of the Pauli field (electron field). Thus, the Pauli field has an electric charge and an electric current continuously distributed in space, which are not reduced to point charged particles and their motion. The electric charge of the Pauli field contained in some region of space $\Omega$ is equal to $q = \int \rho dV$, where the integral is taken over the region $\Omega$. In particular, the electric charge of the electron field of an electrically neutral atom whose nucleus has the charge $Ze$ is equal to $\int \rho dV = -Ze$, where $Z = 1,2,...$, while the integral is taken over the entire space. Taking into account (22), we come to the conclusion that the wave function in an electrically neutral atom satisfies the normalization condition

$$\int \Psi^*\Psi dV = Z \qquad (25)$$

For a hydrogen atom $Z = 1$.

As follows from condition (25), it is assumed here that equations (15)–(23) are valid not only for the hydrogen atom, but also for any other atoms and, in general, for any electron field under any conditions. This is the fundamental difference between equation (15) and the usual Pauli equation (and the Schrödinger equation), which is valid only for a "single-electron" system. In quantum mechanics, in order to describe an electron system that has an electric charge $-Ze$, where $Z = 2,3, ...$, for example, the multi-electron Schrödinger equation is used, written in a $3Z$-dimensional configuration space. We assert, and in the future we will try to prove it, that equations (15)-(23), written in the usual 3-dimensional physical space, are able to describe any system consisting of electromagnetic and electron fields, regardless of the total charge of the electron field, including atoms with any atomic number $Z$.

Direct calculation gives

$$\Psi^*\boldsymbol{\sigma}\Psi = (\psi_1^*\psi_2 + \psi_1\psi_2^*, i(\psi_1\psi_2^* - \psi_1^*\psi_2), |\psi_1|^2 - |\psi_2|^2) \tag{26}$$

In the general case, using relation (26), one obtains

$$|\Psi^*\boldsymbol{\sigma}\Psi|^2 = (\Psi^*\Psi)^2 \tag{27}$$

This means that the vector $\Psi^*\boldsymbol{\sigma}\Psi$ can always be represented as

$$\Psi^*\boldsymbol{\sigma}\Psi = \boldsymbol{\nu}\Psi^*\Psi \tag{28}$$

where $\boldsymbol{\nu}$ is a unit vector ($\boldsymbol{\nu}^2 = 1$), which in the general case can depend on time and spatial coordinates.

Note that the system of equations (15)-(23) is not relativistically invariant, since equations (15) and (16) do not have relativistic invariance. A relativistic-invariant theory can be constructed on the basis of the Dirac equation. The next paper will be devoted to this issue.

Let us show that equation (15) under certain conditions transforms into the conventional linear Schrödinger equation.

The solution of equation (16) for a localized electron field in an infinite empty ($\mathbf{G} = 0$) space has the form [29]

$$\mathbf{G}_\Sigma = \mathbf{G}_e = e \int \frac{(\Psi^*\boldsymbol{\sigma}\Psi)_{t-R/c}}{R} dV' \tag{29}$$

Taking into account (26), it is easy to obtain by direct calculation

$$\boldsymbol{\sigma}\mathbf{G}_e\Psi = e \begin{pmatrix} 2\psi_2 \int \frac{(\psi_1\psi_2^*)_{t-R/c}}{R} dV' \\ 2\psi_1 \int \frac{(\psi_1^*\psi_2)_{t-R/c}}{R} dV' \end{pmatrix} + e \begin{pmatrix} \psi_1 \\ -\psi_2 \end{pmatrix} \int \frac{|\psi_1|^2_{t-R/c} - |\psi_2|^2_{t-R/c}}{R} dV' \tag{30}$$

Then, taking into account (9), (12), and (30), we obtain

$$(e\varphi_e + e\boldsymbol{\sigma}\mathbf{G}_e)\Psi = \begin{pmatrix} -2e^2\psi_1 \int \frac{|\psi_2|^2_{t-R/c}}{R} dV' + 2e^2\psi_2 \int \frac{(\psi_1\psi_2^*)_{t-R/c}}{R} dV' \\ -2e^2\psi_2 \int \frac{|\psi_1|^2_{t-R/c}}{R} dV' + 2e^2\psi_1 \int \frac{(\psi_1^*\psi_2)_{t-R/c}}{R} dV' \end{pmatrix} \tag{31}$$

In what follows, for brevity, the components of the spinor $\Psi$ will be called polarizations of the electron wave. Thus, in the Pauli approximation, the electron wave has two polarizations.

Taking into account (31), we write Eq. (15) in explicit form for each polarization of the electron wave:

$$i\hbar \frac{\partial \psi_1}{\partial t} = \frac{1}{2m_e}\left(\frac{\hbar}{i}\nabla + \frac{e}{c}\mathbf{A}_\Sigma\right)^2 \psi_1 + \left(-e\varphi + 2e^2 \int \frac{|\psi_2|^2_{t-R/c}}{R}dV' + \mu_B H_{\Sigma z} - eG_z\right)\psi_1 -$$

$$\left(2e^2 \int \frac{(\psi_1\psi_2^*)_{t-R/c}}{R}dV' - \mu_B(H_{\Sigma x} - iH_{\Sigma y}) + e(G_x - iG_y)\right)\psi_2 \qquad (32)$$

$$i\hbar \frac{\partial \psi_2}{\partial t} = \frac{1}{2m_e}\left(\frac{\hbar}{i}\nabla + \frac{e}{c}\mathbf{A}_\Sigma\right)^2 \psi_2 + \left(-e\varphi + 2e^2 \int \frac{|\psi_1|^2_{t-R/c}}{R}dV' - \mu_B H_{\Sigma z} + eG_z\right)\psi_2 -$$

$$\left(2e^2 \int \frac{(\psi_1^*\psi_2)_{t-R/c}}{R}dV' - \mu_B(H_{\Sigma x} + iH_{\Sigma y}) + e(G_x - iG_y)\right)\psi_1 \qquad (33)$$

where $\mathbf{H}_\Sigma = (H_{\Sigma x}, H_{\Sigma y}, H_{\Sigma z})$; $\mathbf{G} = (G_x, G_y, G_z)$; $\mu_B = \frac{e\hbar}{2m_e c}$ is the Bohr magneton.

Equations (32) and (33) are similar in its structure to the Hartree-Fock equations. This gives hope that the theory (15) - (23) is applicable, including for "multi-electron" atoms ($Z > 1$). This issue will be discussed in future papers in this series.

In the case when $\mathbf{G} = 0$ and the last term in equation (15) can be neglected, equations (32) and (33) have solutions

$$\Psi = \begin{pmatrix} a_1\psi \\ a_2\psi \end{pmatrix} \qquad (34)$$

where $a_1$ and $a_2$ are arbitrary constants; the function $\psi$ satisfies the Schrödinger equation

$$i\hbar \frac{\partial \psi}{\partial t} = \left[\frac{1}{2m_e}\left(\frac{\hbar}{i}\nabla + \frac{e}{c}\mathbf{A}_\Sigma\right)^2 - e\varphi\right]\psi \qquad (35)$$

while the vector potential $\mathbf{A}_\Sigma$ is determined by relation (10) and includes both the external electromagnetic field and the own electromagnetic field of the electron wave. This is the fundamental difference between equation (35) and the conventional linear Schrödinger equation, in which the vector potential refers only to an external electromagnetic field.

Taking into account the normalization condition (25), we conclude that the constants $a_1$ and $a_2$ characterize the distribution of the electric charge of an electron wave between its polarizations. In particular, for the hydrogen atom, in which the function $\psi$ is normalized to unity, i.e., $\int |\psi|^2 dV = 1$, the constants $a_1$ and $a_2$ satisfy the normalization condition

$$|a_1|^2 + |a_2|^2 = 1 \qquad (36)$$

which has a simple physical meaning [22-24]: the total electric charge of an electron wave in a hydrogen atom, regardless of how it is distributed over polarizations, is equal to $-e$.

Equation (35) coincides with equation (2) obtained in [22]. It has the necessary properties: it does not contain the own electrostatic field of an electron wave, but it contains own radiative electromagnetic field of the electron wave that arises in an excited atom, in particular, during spontaneous emission. As shown in [22-26], equation (35) gives the correct spontaneous emission spectrum of a hydrogen atom, and at the same time allows describing all the basic quantum effects in the framework of classical field theory without field quantization. Thus, we come to the conclusion that, at least in the Schrödinger approximation, the self-consistent system of equations (15)–(23) also allows describing all basic quantum effects within the framework of classical field theory. We will show this next.

## 3. Conservation laws for the Maxwell-Pauli field

### 3.1. Charge conservation law

First of all, from equation (15), which describes the charged field $\Psi$, the law of conservation of the electric charge of the electron wave follows:

$$\frac{\partial \rho}{\partial t} + \text{div } \mathbf{j} = 0 \tag{37}$$

It is easy to show that separately the parameters $|\psi_1|^2$ and $|\psi_2|^2$ do not satisfy the continuity equations, i.e. electric charges of the electron field belonging to different polarizations are not conserved; only the total charge of the entire electron field is conserved. This means that the electron field, together with its charge, can be redistributed between polarizations. Thus, the processes are possible in which the electric charge of the electron field flow from one polarization to another, creating internal (for example, intra-atomic) currents.

Formally, the continuity equation (37) also follows from Maxwell equations (19) and (20). On this basis, it is sometimes argued that the law of conservation of electric charge follows from Maxwell equations. In reality, this only speaks of the consistency of Maxwell equations with the law of conservation of charge, which is a property of charged matter (for example, an electron field), and the statement that the law of conservation of charge follows from Maxwell equations is incorrect, because for Maxwell equations, charges and currents are external objects and obey their own equations (in the case under consideration, the Pauli equation (15)).

### 3.2. Magnetic moment of the Pauli field

As we have seen, the Pauli field described by equation (15) is a classical (continuously distributed in space) electrically charged matter. The redistribution of the electrical charge of the Pauli field in space is described by the current density $\mathbf{j}$. Like any classical electrically charged matter continuously distributed in space, the Pauli field has a magnetic moment [29]

$$\mathbf{M} = \frac{1}{2c} \int \mathbf{r} \times \mathbf{j} dV \tag{38}$$

We write the electric current density (23) as

$$\mathbf{j} = \mathbf{j}_0 + c\,\text{rot}\,\mathfrak{m} \tag{39}$$

where

$$\mathbf{j}_0 = \frac{e\hbar}{2m_e i}\left[(\nabla\Psi^*)\Psi - \Psi^*\nabla\Psi\right] - \frac{e^2}{m_e c}\mathbf{A}_\Sigma \Psi^*\Psi \tag{40}$$

$$\mathfrak{m} = -\mu_B \Psi^* \boldsymbol{\sigma}\Psi \tag{41}$$

Substituting (39) into relation (38), one obtains

$$\mathbf{M} = \frac{1}{2c}\int \mathbf{r}\times\mathbf{j}_0 dV + \frac{1}{2}\int \mathbf{r}\times\text{rot}\,\mathfrak{m}\,dV \tag{42}$$

Using well-known vector identities, it is easy to obtain

$$\mathbf{r}\times\text{rot}\,\mathfrak{m} = \nabla(\mathbf{r}\mathfrak{m}) - \frac{\partial}{\partial x_i}(x_i \mathfrak{m}) + 2\mathfrak{m} \tag{43}$$

We substitute (43) into (42), taking into account that, according to the Gauss theorem, the volume integrals of the first two terms on the right-hand side of (43) turn into surface ones. Then for a localized electron field (for example, in an atom), for which the vector $\mathfrak{m}$ rapidly decreases at infinity, we obtain

$$\mathbf{M} = \mathbf{M}_{or} + \boldsymbol{\mu} \tag{44}$$

where

$$\mathbf{M}_{or} = \frac{1}{2c}\int \mathbf{r}\times\mathbf{j}_0 dV \tag{45}$$

$$\boldsymbol{\mu} = \int \mathfrak{m}\,dV = -\mu_B \int \Psi^*\boldsymbol{\sigma}\Psi\,dV \tag{46}$$

Thus, the magnetic moment of a localized electron field has two components: the component $\mathbf{M}_{or}$, which depends on the choice of the point relative to which the magnetic moment is calculated, and the component $\boldsymbol{\mu}$, which does not depend on this choice. If we formally liken the Pauli field to a classical electrically charged magnetic fluid, then we come to the conclusion that the current density $\mathbf{j}_0$ and the component of the magnetic moment $\mathbf{M}_{or}$ are related to the "flow" of electrically charged matter, i.e. with convective currents. Therefore, in the general case, $\mathbf{M}_{or}$ can be called the convective component of the magnetic moment of the electron field. If the magnetic moment of the electron field in an atom is calculated with respect to the nucleus of the atom, then according to the tradition that originates in Bohr's naive theory, the component $\mathbf{M}_{or}$ is called the orbital magnetic moment. At the same time, according to (46), the component of the magnetic moment $\boldsymbol{\mu}$ is the intrinsic magnetic moment of the Pauli field, continuously distributed in space with a density $\mathfrak{m}$, which is an internal characteristic of the electron field at each of its points. The distributed intrinsic magnetic moment $\mathfrak{m}$ of the electron field is similar to the internal magnetic moment of a classical magnetic fluid and is not related to convective currents. Thus, we come to the conclusion that the Pauli electron field has its own magnetic moment, continuously distributed in space with density $\mathfrak{m}$ [21].

From this, the meaning of decomposition (39) of the current density of an electron wave into two components becomes clear: component $\mathbf{j}_0$ describes the convective currents of an electron wave (i.e., currents associated with the flow of an electrically charged continuous medium), while component $c\operatorname{rot}\mathbf{m}$ describes the internal currents of a magnetized continuous medium (i.e. an electron wave), which is not associated with the convective motion of charged matter.

### 3.3. Energy-momentum conservation law

To obtain other conservation laws, consider the functional

$$S = \frac{1}{c}\int \mathcal{L}\,d\Omega \tag{47}$$

where $d\Omega = cdtdV$;

$$\mathcal{L} = \frac{i\hbar}{2}\left(\Psi^*\frac{\partial\Psi}{\partial t} - \frac{\partial\Psi^*}{\partial t}\Psi\right) + \frac{1}{2m_e}\left(\frac{\hbar}{i}\nabla\Psi^* - \frac{e}{c}\mathbf{A}_\Sigma\Psi^*\right)\left(\frac{\hbar}{i}\nabla\Psi + \frac{e}{c}\mathbf{A}_\Sigma\Psi\right) + e\Psi^*\Psi\varphi_\Sigma + e\Psi^*\boldsymbol{\sigma}\Psi\mathbf{G}_\Sigma - \frac{e\hbar}{2m_e c}\mathbf{A}_\Sigma\operatorname{rot}(\Psi^*\boldsymbol{\sigma}\Psi) + \frac{1}{8\pi}(\mathbf{E}_\Sigma^2 - \mathbf{H}_\Sigma^2) + \frac{1}{8\pi c^2}\frac{\partial\mathbf{G}_\Sigma}{\partial t}\frac{\partial\mathbf{G}_\Sigma}{\partial t} - \frac{1}{8\pi}\frac{\partial\mathbf{G}_\Sigma}{\partial x_k}\frac{\partial\mathbf{G}_\Sigma}{\partial x_k} \tag{48}$$

Obviously, function (48) is real:

$$\mathcal{L}^* = \mathcal{L}$$

It is easy to see directly that the system of equations (15)–(23) follows from the variational principle

$$\delta S = 0 \tag{49}$$

if the functions $\Psi, \Psi^*, \mathbf{G}_\Sigma, \varphi_\Sigma$ and $\mathbf{A}_\Sigma$ are considered independent when varying.

Indeed, varying (47), (48) with respect to $\Psi^*$, taking into account (49), we obtain the Pauli equation (15); variation with respect to $\Psi$ leads to the complex conjugate Pauli equation

$$-i\hbar\frac{\partial\Psi^*}{\partial t} = \frac{1}{2m_e}\left(-\frac{\hbar}{i}\nabla + \frac{e}{c}\mathbf{A}_\Sigma\right)^2\Psi^* - e\varphi_\Sigma\Psi^* - e\Psi^*\boldsymbol{\sigma}\mathbf{G}_\Sigma + \frac{e\hbar}{2m_e c}\Psi^*\boldsymbol{\sigma}\mathbf{H}_\Sigma \tag{50}$$

variation with respect to $\mathbf{G}_\Sigma$ leads to equation (16), while variation with respect to $\varphi_\Sigma$ and $\mathbf{A}_\Sigma$, taking into account (21), leads to Maxwell equations (19) and (20).

Since the theory under consideration is not relativistically invariant, we cannot speak about the stress-energy tensor of the Maxwell-Pauli field. However, we can determine the energy density and momentum density of this field.

In the general case, the energy density of the electron field must contain a component $m_e c^2 \Psi^*\Psi$ corresponding to the rest energy of this field. It is easy to see that Lagrangian (48) does not have a component corresponding to the rest energy of the electron wave. In order for such a component to appear explicitly, we introduce an auxiliary spinor

$$\Phi = \Psi \exp(-im_e c^2 t/\hbar) \tag{51}$$

Then we rewrite Lagrangian (48) in the form

$$\mathcal{L} = -m_e c^2 \Phi^* \Phi + \frac{i\hbar}{2}\left(\Phi^* \frac{\partial \Phi}{\partial t} - \frac{\partial \Phi^*}{\partial t}\Phi\right) + \frac{1}{2m_e}\left(\frac{\hbar}{i}\nabla\Phi^* - \frac{e}{c}\mathbf{A}_\Sigma \Phi^*\right)\left(\frac{\hbar}{i}\nabla\Phi + \frac{e}{c}\mathbf{A}_\Sigma \Phi\right) + e\Phi^*\Phi\varphi_\Sigma +$$
$$e\Phi^*\boldsymbol{\sigma}\Phi\mathbf{G}_\Sigma - \frac{e\hbar}{2m_e c}\mathbf{A}_\Sigma \mathrm{rot}(\Phi^*\boldsymbol{\sigma}\Phi) + \frac{1}{8\pi}(\mathbf{E}_\Sigma^2 - \mathbf{H}_\Sigma^2) + \frac{1}{8\pi c^2}\frac{\partial \mathbf{G}_\Sigma}{\partial t}\frac{\partial \mathbf{G}_\Sigma}{\partial t} - \frac{1}{8\pi}\frac{\partial \mathbf{G}_\Sigma}{\partial x_k}\frac{\partial \mathbf{G}_\Sigma}{\partial x_k} \tag{52}$$

In this case, the Pauli field is described by the spinor $\Phi$, and the Maxwell-Pauli field equations are obtained from (47), (49) and (52), with independent variation of $\Phi, \Phi^*, \mathbf{G}_\Sigma, \varphi_\Sigma$ and $\mathbf{A}_\Sigma$. It is easy to see that the transition from Lagrangian (48) to Lagrangian (52) leads to change only equation (15), and leaves equations (16)-(23) unchanged. In this case, the Pauli equation for spinor $\Phi$ differs from equation (15) by an additional term $m_e c^2 \Psi$, which appears on the right-hand side. Substituting relation (51) into this equation brings it to the form (15).

When calculating the energy density and momentum density of the Maxwell-Pauli field, it is necessary to use the Lagrangian (52).

According to the general rule for calculating the energy density and momentum density of a classical field [29], one obtains:

$$W = \frac{\partial \mathcal{L}}{\partial \Phi_{,t}}\Phi_{,t} + \Phi^*_{,t}\frac{\partial \mathcal{L}}{\partial \Phi^*_{,t}} + \mathbf{G}_{\Sigma,t}\frac{\partial \mathcal{L}}{\partial \mathbf{G}_{\Sigma,t}} + \varphi_{\Sigma,t}\frac{\partial \mathcal{L}}{\partial \varphi_{\Sigma,t}} + \mathbf{A}_{\Sigma,t}\frac{\partial \mathcal{L}}{\partial \mathbf{A}_{\Sigma,t}} - \mathcal{L} \tag{53}$$

is the energy density,

$$J_k = \frac{\partial \mathcal{L}}{\partial \Phi_{,k}}\Phi_{,t} + \Phi^*_{,t}\frac{\partial \mathcal{L}}{\partial \Phi^*_{,k}} + \mathbf{G}_{\Sigma,t}\frac{\partial \mathcal{L}}{\partial \mathbf{G}_{\Sigma,k}} + \varphi_{\Sigma,t}\frac{\partial \mathcal{L}}{\partial \varphi_{\Sigma,k}} + \mathbf{A}_{\Sigma,t}\frac{\partial \mathcal{L}}{\partial \mathbf{A}_{\Sigma,k}} \tag{54}$$

is the energy flux density,

$$P_k = -\left(\frac{\partial \mathcal{L}}{\partial \Phi_{,t}}\Phi_{,k} + \Phi^*_{,k}\frac{\partial \mathcal{L}}{\partial \Phi^*_{,t}} + \mathbf{G}_{\Sigma,k}\frac{\partial \mathcal{L}}{\partial \mathbf{G}_{\Sigma,t}} + \varphi_{\Sigma,k}\frac{\partial \mathcal{L}}{\partial \varphi_{\Sigma,t}} + \mathbf{A}_{\Sigma,k}\frac{\partial \mathcal{L}}{\partial \mathbf{A}_{\Sigma,t}}\right) \tag{55}$$

is the momentum density,

$$\sigma_{ik} = -\left(\frac{\partial \mathcal{L}}{\partial \Phi_{,k}}\Phi_{,i} + \Phi^*_{,i}\frac{\partial \mathcal{L}}{\partial \Phi^*_{,k}} + \mathbf{G}_{\Sigma,i}\frac{\partial \mathcal{L}}{\partial \mathbf{G}_{\Sigma,k}} + \varphi_{\Sigma,i}\frac{\partial \mathcal{L}}{\partial \varphi_{\Sigma,k}} + \mathbf{A}_{\Sigma,i}\frac{\partial \mathcal{L}}{\partial \mathbf{A}_{\Sigma,k}} - \delta_{ik}\mathcal{L}\right) \tag{56}$$

is the momentum flux density (three-dimensional stress tensor), where $\Phi_{,t} \equiv \frac{\partial \Phi}{\partial t}$; $\Phi_{,k} \equiv \frac{\partial \Phi}{\partial x_k}$, etc.

Note that in this paper, considering the nonrelativistic theory, we do not distinguish between the covariant and contravariant components of vectors. All vector indices run through the values 1,2,3, and summation is performed over repeated indices.

By definition, parameters (53)-(56) satisfy the continuity equations [29]

$$\frac{\partial W}{\partial t} + \mathrm{div}\mathbf{J} = 0 \tag{57}$$

$$\frac{\partial P_i}{\partial t} + \frac{\partial \sigma_{ik}}{\partial x_k} = 0 \tag{58}$$

Equation (57) expresses the energy conservation law, while equation (58) expresses the momentum conservation law for the Maxwell-Pauli field.

Carrying out calculations taking into account (20), (21)-(23), and passing from spinor $\Phi$ to spinor $\Psi$, one obtains (some details of the derivation are given in the Appendix)

$$W = -\frac{m_e c^2}{e}\rho + W' \tag{59}$$

$$\mathbf{J} = -\frac{m_e c^2}{e}\mathbf{j} + \mathbf{J}' \tag{60}$$

$$\mathbf{P} = \frac{1}{4\pi c}\mathbf{E}_\Sigma \times \mathbf{H}_\Sigma - \frac{m_e}{e}\mathbf{j}_0 - \frac{1}{4\pi c^2}\frac{\partial G_{\Sigma k}}{\partial t}\nabla G_{\Sigma k} \tag{61}$$

$$\sigma_{ik} = -\frac{1}{4\pi}E_{\Sigma i}E_{\Sigma k} - \frac{1}{4\pi}H_{\Sigma i}H_{\Sigma k} + \frac{\hbar^2}{2m_e}\left(\frac{\partial \Psi^*}{\partial x_k}\frac{\partial \Psi}{\partial x_i} + \frac{\partial \Psi^*}{\partial x_i}\frac{\partial \Psi}{\partial x_k}\right) - \frac{e\hbar}{i2m_e c}A_{\Sigma i}\left(\frac{\partial \Psi^*}{\partial x_k}\Psi - \Psi^*\frac{\partial \Psi}{\partial x_k}\right) -$$

$$\frac{\hbar e}{i2m_e c}A_{\Sigma k}\left(\frac{\partial \Psi^*}{\partial x_i}\Psi - \Psi^*\frac{\partial \Psi}{\partial x_i}\right) + \frac{e^2}{m_e c^2}A_{\Sigma i}A_{\Sigma k}\Psi^*\Psi + \frac{1}{4\pi}\frac{\partial \mathbf{G}_\Sigma}{\partial x_i}\frac{\partial \mathbf{G}_\Sigma}{\partial x_k} + \delta_{ik}\left(\frac{1}{8\pi}(\mathbf{E}_\Sigma^2 + \mathbf{H}_\Sigma^2) + \right.$$

$$\frac{i\hbar}{2}\left(\Psi^*\frac{\partial \Psi}{\partial t} - \frac{\partial \Psi^*}{\partial t}\Psi\right) + \frac{1}{2m_e}\left(\frac{\hbar}{i}\nabla\Psi^* - \frac{e}{c}\mathbf{A}_\Sigma\Psi^*\right)\left(\frac{\hbar}{i}\nabla\Psi + \frac{e}{c}\mathbf{A}_\Sigma\Psi\right) + e\Psi^*\Psi\varphi_\Sigma + e\Psi^*\boldsymbol{\sigma}\Psi\mathbf{G}_\Sigma -$$

$$\left.\frac{e\hbar}{2m_e c}\mathbf{A}_\Sigma\mathrm{rot}(\Psi^*\boldsymbol{\sigma}\Psi) + \frac{1}{8\pi c^2}\frac{\partial \mathbf{G}_\Sigma}{\partial t}\frac{\partial \mathbf{G}_\Sigma}{\partial t} - \frac{1}{8\pi}\frac{\partial \mathbf{G}_\Sigma}{\partial x_j}\frac{\partial \mathbf{G}_\Sigma}{\partial x_j}\right) + \frac{e\hbar}{2m_e c}(\delta_{il}\varepsilon_{skm} + \delta_{is}\varepsilon_{klm})A_{\Sigma s}\frac{\partial \Psi^*\sigma_m\Psi}{\partial x_l} \tag{62}$$

where the components $-\frac{m_e c^2}{e}\rho$ and $-\frac{m_e c^2}{e}\mathbf{j}$ describe the rest energy density and the rest energy flux density of the electron field, while

$$W' = \frac{1}{8\pi}\left(\mathbf{E}_\Sigma^2 + \mathbf{H}_\Sigma^2\right) + \frac{1}{2m_e}\left(-\frac{\hbar}{i}\nabla\Psi^* + \frac{e}{c}\mathbf{A}_\Sigma\Psi^*\right)\left(\frac{\hbar}{i}\nabla\Psi + \frac{e}{c}\mathbf{A}_\Sigma\Psi\right) - \Psi^*\boldsymbol{\sigma}\Psi(-\mu_B\mathbf{H}_\Sigma + e\mathbf{G}_\Sigma) +$$

$$\frac{1}{8\pi}\left(\frac{1}{c^2}\frac{\partial \mathbf{G}_\Sigma}{\partial t}\frac{\partial \mathbf{G}_\Sigma}{\partial t} + \frac{\partial \mathbf{G}_\Sigma}{\partial x_k}\frac{\partial \mathbf{G}_\Sigma}{\partial x_k}\right) \tag{63}$$

and

$$\mathbf{J}' = \frac{c}{4\pi}\mathbf{E}_\Sigma \times \mathbf{H}_\Sigma - \frac{\hbar}{2m_e i}\left(-\frac{\hbar}{i}\nabla\Psi^* + \frac{e}{c}\mathbf{A}_\Sigma\Psi^*\right)\frac{\partial \Psi}{\partial t} + \frac{\hbar}{2m_e i}\frac{\partial \Psi^*}{\partial t}\left(\frac{\hbar}{i}\nabla\Psi + \frac{e}{c}\mathbf{A}_\Sigma\Psi\right) - \mu_B\frac{\partial \mathbf{A}_\Sigma}{\partial t}\times\Psi^*\boldsymbol{\sigma}\Psi -$$

$$\frac{1}{4\pi}\frac{\partial G_{\Sigma k}}{\partial t}\nabla G_{\Sigma k} - \varphi_\Sigma \mathbf{j} + \frac{c}{4\pi}\mathrm{rot}(\varphi_\Sigma \mathbf{H}_\Sigma) \tag{64}$$

are the non-relativistic components of the energy density and energy flux density.

Taking into account Eq. (37), it can be seen that the nonrelativistic components of the energy density (63) and the energy flux density (64) also satisfy the continuity equation (57), i.e., the nonrelativistic law of conservation of energy holds.

The components included in the energy density (63) are easy to identify:

$$W_{\mathbf{EH}} = \frac{1}{8\pi}\left(\mathbf{E}_\Sigma^2 + \mathbf{H}_\Sigma^2\right) \tag{65}$$

is the energy density of the classical electromagnetic field [29];

$$W'_\Psi = \frac{1}{2m_e}\left(-\frac{\hbar}{i}\nabla\Psi^* + \frac{e}{c}\mathbf{A}_\Sigma\Psi^*\right)\left(\frac{\hbar}{i}\nabla\Psi + \frac{e}{c}\mathbf{A}_\Sigma\Psi\right) \tag{66}$$

is the energy density of the electron field (Pauli field), associated with the convective motion of field matter;

$$W_\mathbf{G} = \frac{1}{8\pi}\left(\frac{1}{c^2}\frac{\partial \mathbf{G}_\Sigma}{\partial t}\frac{\partial \mathbf{G}_\Sigma}{\partial t} + \frac{\partial \mathbf{G}_\Sigma}{\partial x_k}\frac{\partial \mathbf{G}_\Sigma}{\partial x_k}\right) \tag{67}$$

is the energy density of the $\mathbf{G}$-field;

$$W_{\boldsymbol{\mu}} = -\Psi^*\boldsymbol{\sigma}\Psi(-\mu_B\mathbf{H}_\Sigma + e\mathbf{G}_\Sigma) \tag{68}$$

is the energy density associated with the interaction of the intrinsic magnetic moment of the electron field with the classical magnetic field and the **G**-field.

Similarly, the components of the energy flux density (64):

$$\mathbf{J_{EH}} = \frac{c}{4\pi} \mathbf{E}_\Sigma \times \mathbf{H}_\Sigma \qquad (69)$$

is the Poynting vector [29] – the energy flux density of the classical electromagnetic field;

$$\mathbf{J'_\Psi} = -\frac{\hbar}{2m_e i}\left(-\frac{\hbar}{i}\nabla\Psi^* + \frac{e}{c}\mathbf{A}_\Sigma\Psi^*\right)\frac{\partial \Psi}{\partial t} + \frac{\hbar}{2m_e i}\frac{\partial \Psi^*}{\partial t}\left(\frac{\hbar}{i}\nabla\Psi + \frac{e}{c}\mathbf{A}_\Sigma\Psi\right) - \varphi_\Sigma \mathbf{j} \qquad (70)$$

is the energy flux density of the electron field (Pauli field) associated with the convective motion of field matter;

$$\mathbf{J_G} = -\frac{1}{4\pi}\frac{\partial G_{\Sigma k}}{\partial t}\nabla G_{\Sigma k} \qquad (71)$$

is the energy flux density of the **G**-field;

$$\mathbf{J_\mu} = -\mu_B \frac{\partial \mathbf{A}_\Sigma}{\partial t} \times \Psi^* \boldsymbol{\sigma} \Psi \qquad (72)$$

is the energy flux density associated with the interaction of the intrinsic magnetic moment of the electron field with the electromagnetic field. Note that all energy flux densities are determined up to the curl of an arbitrary vector.

Using the Maxwell equations, one obtains the energy conservation law for the classical electromagnetic field [29]

$$\frac{\partial W_{EH}}{\partial t} + \text{div}\mathbf{J_{EH}} = -\mathbf{jE} \qquad (73)$$

where the term on the right-hand side describes the rate of change in the energy of the electromagnetic field due to its interaction with electric currents.

Similarly, using equation (16) and relations (67) and (71), one obtains the energy conservation law for the **G**-field

$$\frac{\partial W_\mathbf{G}}{\partial t} + \text{div}\mathbf{J_G} = e(\Psi^*\boldsymbol{\sigma}\Psi)\frac{\partial \mathbf{G}_\Sigma}{\partial t} \qquad (74)$$

where the term on the right-hand side describes the rate of change in the energy of the **G**-field due to its interaction with the intrinsic magnetic moment of the electron field.

The presence of the rest energy of the electron field (term $-\frac{m_e c^2}{e}\rho$ in equation (59)) means that for the electron field one can introduce (at least formally) a rest mass, which is continuously distributed in space with a density $\rho_m = -\frac{m_e}{e}\rho$.

Term $\mathbf{j}_{m0} = -\frac{m_e}{e}\mathbf{j}_0$ in equation (61) describes the electron field momentum density associated with the convective motion of the electron field matter. As for any continuous medium, for the electron field, one can introduce the speed of the flow (convective motion) of matter

$$\mathbf{v}_m = \frac{\mathbf{j}_{m0}}{\rho_m} \qquad (75)$$

and the speed of the flow (convective transfer) of electric charge

$$\mathbf{v}_e = \frac{\mathbf{j}_0}{\rho} \tag{76}$$

It can be seen that the matter (mass) and the electric charge of the electron wave are transferred at the same speed

$$\mathbf{v} = \mathbf{v}_m = \mathbf{v}_e \tag{77}$$

i.e., inextricably linked.

### 3.4. Convective angular momentum of the Maxwell-Pauli field

The Maxwell-Pauli field described by equations (15)-(23) is considered by us as a classical field, i.e. is a matter continuously distributed in space and time, having energy and momentum continuously distributed in space. Like any classical matter distributed in space, the Maxwell-Pauli field has an angular momentum, which, according to the classical field theory [29], is determined by the relation

$$\mathbf{L} = \int \mathbf{r} \times \mathbf{P} dV \tag{78}$$

We calculate (78) using equation (61). As a result, one obtains

$$\mathbf{L} = \mathbf{L}_{EH} + \mathbf{L}_G + \mathbf{L}_{or} \tag{79}$$

where

$$\mathbf{L}_{EH} = \frac{1}{4\pi c} \int \mathbf{r} \times \mathbf{E}_\Sigma \times \mathbf{H}_\Sigma dV \tag{80}$$

$$\mathbf{L}_G = -\frac{1}{4\pi c^2} \int \mathbf{r} \times \frac{\partial G_{\Sigma k}}{\partial t} \nabla G_{\Sigma k} dV \tag{81}$$

$$\mathbf{L}_{or} = -\frac{m_e}{e} \int \mathbf{r} \times \mathbf{j}_0 dV \tag{82}$$

The angular momentum (79) and all its components (80)-(82) depend on the point relative to which they are calculated. Therefore, they are the convective components of the angular momentum. Component $\mathbf{L}_{EH}$ describes the angular momentum of the classical electromagnetic field, which has a convective nature, i.e., associated with the flow of electromagnetic field matter; component $\mathbf{L}_G$ describes the convective component of the angular momentum of the vector field $\mathbf{G}_\Sigma$; component $\mathbf{L}_{or}$ describes the convective component of the angular momentum of the electron field (Pauli field). If the angular momentum of the electron field in an atom is calculated with respect to the nucleus of the atom, then according to the tradition that originates in the naive Bohr's theory, the component $\mathbf{L}_{or}$ is usually called the orbital angular momentum.

Multiplying Eq. (58) vectorially by $\mathbf{r}$ on the left, one obtains

$$\frac{\partial l_i}{\partial t} + \frac{\partial Q_{ik}}{\partial x_k} = \frac{1}{2}\varepsilon_{ikl}(\sigma_{lk} - \sigma_{kl}) \tag{83}$$

where

$$\mathbf{l} = \mathbf{r} \times \mathbf{P} \tag{84}$$

is the density of angular momentum of Maxwell-Pauli field;

$$Q_{ik} = \varepsilon_{iml} x_m \sigma_{lk} \quad (85)$$

is the flux density of the Maxwell-Pauli field angular momentum.

Using (62), equation (83) can be reduced to the form (see Appendix)

$$\frac{\partial l_i}{\partial t} + \frac{\partial}{\partial x_k}\left(Q_{ik} - \frac{e\hbar}{2m_e c}(A_{\Sigma i}\Psi^*\sigma_k\Psi - \delta_{ik}A_{\Sigma s}\Psi^*\sigma_s\Psi)\right) = \mu_B\big((\Psi^*\boldsymbol{\sigma}\Psi) \times \mathbf{H}_\Sigma\big)_i \quad (86)$$

Taking into account that there is a nondivergent term on the right-hand side of (86), we conclude that the convective angular momentum of the Maxwell-Pauli field (78) is not conserved.

Integrating (86) over the volume, and taking into account (78) and (84), one obtains the equation for the convective angular momentum of the Maxwell-Pauli field

$$\frac{d\mathbf{L}}{dt} = \mu_B \int (\Psi^*\boldsymbol{\sigma}\Psi) \times \mathbf{H}_\Sigma dV = -\int \mathfrak{m} \times \mathbf{H}_\Sigma dV \quad (87)$$

Here it is assumed that the flux of convective angular momentum through an infinite surface is zero.

According to (87), the convective angular momentum of the Maxwell-Pauli field (78) can change only due to the interaction of the intrinsic magnetic moment with the magnetic field.

### 3.5. Intrinsic angular momentum of the Pauli field

Consider the vector

$$\mathbf{s} = \frac{\hbar}{2}\Psi^*\boldsymbol{\sigma}\Psi \quad (88)$$

which characterizes the electron field (Pauli field).

Using the Pauli equation (15) and taking into account the properties of the Pauli matrices [28]

$$(\boldsymbol{\sigma}\mathbf{G})\boldsymbol{\sigma} = \mathbf{G} + i\boldsymbol{\sigma} \times \mathbf{G}, \qquad \boldsymbol{\sigma}(\boldsymbol{\sigma}\mathbf{G}) = \mathbf{G} + i\mathbf{G} \times \boldsymbol{\sigma}$$

for the vector (88) one obtains the equation (see Appendix)

$$\frac{\partial s_i}{\partial t} + \frac{\partial g_{ik}}{\partial x_k} = K_i \quad (89)$$

where

$$g_{ik} = \frac{\hbar}{2m_e}\left[\frac{\hbar}{2i}\left(\Psi^*\sigma_i\frac{\partial \Psi}{\partial x_k} - \frac{\partial \Psi^*}{\partial x_k}\sigma_i\Psi\right) + \frac{e}{c}A_{\Sigma k}\Psi^*\sigma_i\Psi\right] \quad (90)$$

$$\mathbf{K} = \Psi^*\boldsymbol{\sigma}\Psi \times (-\mu_B \mathbf{H}_\Sigma + e\mathbf{G}_\Sigma) \quad (91)$$

or

$$\mathbf{K} = \mathfrak{m} \times \left(\mathbf{H}_\Sigma - \frac{e}{\mu_B}\mathbf{G}_\Sigma\right) \quad (92)$$

Equation (89) describes the law of conservation of the intrinsic angular momentum of the electron field, where $g_{kj}$ is the flux density of the intrinsic angular momentum of the electron field, while vector **K** is the density of the moment of forces acting on the electron field. It follows from equation (92) that in addition to the usual classical moment of forces with a density 𝔪 × **H**$_\Sigma$, which acts on a magnetic moment in a magnetic field, an additional moment of forces distributed in space with a density $e\Psi^*\boldsymbol{\sigma}\Psi \times \mathbf{G}_\Sigma$ acts on an electron wave.

We introduce a vector that characterizes the electron field as a whole (the Pauli field),

$$\mathbf{S} = \frac{\hbar}{2}\int \Psi^*\boldsymbol{\sigma}\Psi dV \tag{93}$$

which is continuously distributed in space with density (88).

As is known, vector (93) is interpreted as intrinsic angular momentum (spin) of a quantum object described by the Pauli equation. Such an interpretation of the vector (93) is not a strict consequence following from some first principles, but follows from the correspondence principle, when it is necessary to find an analogue from classical mechanics for each physical quantity attributed to a quantum object. As will be shown below, the theory under consideration allows rigorously establishing the physical meaning of the vector **S** without resorting to the correspondence principle.

### 3.6. Intrinsic angular momentum of the G-field

Multiply equation (16) on the right vectorially by $\mathbf{G}_\Sigma$. As a result, one obtains

$$\frac{1}{c^2}\frac{\partial^2 \mathbf{G}_\Sigma}{\partial t^2}\times \mathbf{G}_\Sigma - (\Delta \mathbf{G}_\Sigma)\times \mathbf{G}_\Sigma = 4\pi e(\Psi^*\boldsymbol{\sigma}\Psi)\times \mathbf{G}_\Sigma$$

or

$$\frac{1}{c^2}\frac{\partial}{\partial t}\left(\frac{\partial \mathbf{G}_\Sigma}{\partial t}\times \mathbf{G}_\Sigma\right) - \frac{\partial}{\partial x_k}\left(\frac{\partial \mathbf{G}_\Sigma}{\partial x_k}\times \mathbf{G}_\Sigma\right) = 4\pi e(\Psi^*\boldsymbol{\sigma}\Psi)\times \mathbf{G}_\Sigma \tag{94}$$

Equation (94) can be written as

$$\frac{\partial \eta_i}{\partial t} + \frac{\partial w_{ik}}{\partial x_k} = N_i \tag{95}$$

where

$$\boldsymbol{\eta} = \frac{1}{4\pi c^2}\mathbf{G}_\Sigma \times \frac{\partial \mathbf{G}_\Sigma}{\partial t} \tag{96}$$

$$w_{ik} = \frac{1}{4\pi}\varepsilon_{ilm}\frac{\partial G_{\Sigma l}}{\partial x_k}G_{\Sigma m} \tag{97}$$

$$\mathbf{N} = -e(\Psi^*\boldsymbol{\sigma}\Psi)\times \mathbf{G}_\Sigma \tag{98}$$

We introduce a vector characterizing the **G**-field,

$$\boldsymbol{\Gamma} = \frac{1}{4\pi c^2}\int \mathbf{G}_\Sigma \times \frac{\partial \mathbf{G}_\Sigma}{\partial t}dV \tag{99}$$

which is continuously distributed in space with density (96).

The physical meaning of the vector $\boldsymbol{\Gamma}$ will be established later.

### 3.7. Total angular momentum of the Maxwell-Pauli field

Adding equations (86), (89) and (95) taking into account (91) and (98), one obtains

$$\frac{\partial(l_i+s_i+\eta_i)}{\partial t} + \frac{\partial}{\partial x_k}\left(Q_{ik} + g_{ik} + w_{ik} - \frac{e\hbar}{2m_e c}(A_{\Sigma i}\Psi^*\sigma_k\Psi - \delta_{ik}A_{\Sigma s}\Psi^*\sigma_s\Psi)\right) = 0 \quad (100)$$

Integrating equation (100) over volume, one obtains

$$\int(l_i + s_i + \eta_i)dV = -\oint\left(Q_{ik} + g_{ik} + w_{ik} - \frac{e\hbar}{2m_e c}(A_{\Sigma i}\Psi^*\sigma_k\Psi - \delta_{ik}A_{\Sigma s}\Psi^*\sigma_s\Psi)\right)df_k \quad (101)$$

where the integral on the right-hand side is taken over the surface that bounds the scope of integration.

From (101) it follows that the vector

$$\mathbf{L}_\Sigma = \int(\mathbf{l} + \mathbf{s} + \mathbf{\eta})dV \quad (102)$$

is conserved.

Taking into account (79), (93), and (99), one obtains

$$\mathbf{L}_\Sigma = \mathbf{L_{EH}} + \mathbf{L_G} + \mathbf{L}_{or} + \mathbf{S} + \mathbf{\Gamma} \quad (103)$$

By meaning (see Section 3.4), the vector (102), (103) is the total angular momentum of the Maxwell-Pauli field.

It can be seen from relations (80)-(82) that the components of the angular momentum $\mathbf{L_{EH}}$, $\mathbf{L_G}$ and $\mathbf{L}_{or}$ depend on the choice of the point relative to which the angular momentum is calculated, while the component $\mathbf{S}$ of the angular momentum of the electron field and the component $\mathbf{\Gamma}$ of the angular momentum of the $\mathbf{G}$-field do not depend on this choice and represent the intrinsic angular momenta (spins) of the electron field (Pauli field) and the $\mathbf{G}$-field.

Note that, according to (103), the electromagnetic field does not have its own angular momentum (spin), but only the convective (orbital) component (80) of the angular momentum. This, it would seem, contradicts the modern ideas of quantum mechanics, that "photons" have a spin equal to 1. However, if we consider that the usual formulation of quantum mechanics does not take into account the $\mathbf{G}$-field, which has its own angular momentum (spin) $\mathbf{\Gamma}$, then we can assume that the spin attributed to the "photon" actually refers to the $\mathbf{G}$-field, which ensures the implementation of the law of conservation of angular momentum in various processes.

Thus, the electron field has its own angular momentum (spin) $\mathbf{S}$, which is continuously distributed in space with density (88).

On the other hand, according to (46), the electron field has its own magnetic moment, continuously distributed in space with density (41).

From (41) and (88), one can obtain the relation

$$\mathfrak{m} = -\frac{e}{m_e c}\mathbf{s} \quad (104)$$

which shows that the internal gyromagnetic ratio of the electron field at all points in space is the same and equal to

$$\gamma_e = -\frac{e}{m_e c} \tag{105}$$

Comparing (46) and (93), one obtains

$$\boldsymbol{\mu} = -\frac{e}{m_e c}\mathbf{S} \tag{106}$$

This relationship shows that the intrinsic magnetic moment of an electron wave is always collinear to its intrinsic angular momentum (spin), while the gyromagnetic ratio is constant and equal to (105).

Taking into account (45) and (82), one obtains

$$\mathbf{M}_{or} = -\frac{e}{2m_e c}\mathbf{L}_{or} \tag{107}$$

Relation (107) establishes a relationship between the convective (orbital) component of the angular momentum and the convective (orbital) component of the magnetic moment of the electron field. It completely coincides with the corresponding relation for classical charged matter (for example, classical particles) [29], in particular, the convective (orbital) gyromagnetic ratio for the electron field is determined by the classical expression

$$\gamma_{or} = -\frac{e}{2m_e c} \tag{108}$$

Comparison of (105) and (108) shows that the spin (intrinsic) gyromagnetic ratio of the electron field (105) is twice the convective (orbital) gyromagnetic ratio (107). In the traditional (corpuscular) interpretation of quantum mechanics, this fact has no explanation, and is considered as a purely quantum property of particles. In the considered classical field theory, this fact is a natural property of the electron field, which is an electrically charged magnetic matter, continuously distributed in space. In particular, such properties of the electronic field as its own angular momentum (spin) and associated its own magnetic moment are properties of the field itself and are not reduced to the motions of any particles.

We see that within the framework of self-consistent classical field theory (15)-(23) the problem of spin and intrinsic magnetic moment has a simple and clear explanation, and does not lead to paradoxes.

Taking into account (89)-(91), (93) and (106), one obtains the equation for the spin magnetic moment of the Pauli field

$$\frac{d\boldsymbol{\mu}}{dt} = -\frac{e}{m_e c}\int\bigl(\Psi^*\boldsymbol{\sigma}\Psi \times (-\mu_B \mathbf{H}_\Sigma + e\mathbf{G}_\Sigma)\bigr)dV \tag{109}$$

### 3.8. Total momentum of the Maxwell-Pauli field

Formally, relation (102) can be written as

$$\mathbf{L}_\Sigma = \int(\mathbf{r}\times\mathbf{P}_\Sigma)dV \tag{110}$$

where taking into account (43),

$$\mathbf{P}_\Sigma = \mathbf{P} + \frac{\hbar}{4}\text{rot}(\Psi^*\boldsymbol{\sigma}\Psi) + \frac{1}{8\pi c^2}\text{rot}\left(\mathbf{G}_\Sigma \times \frac{\partial \mathbf{G}_\Sigma}{\partial t}\right) \qquad (111)$$

is the total momentum density of the Maxwell-Pauli field.

Taking into account (61), one writes (111) as

$$\mathbf{P}_\Sigma = \frac{1}{4\pi c}\mathbf{E}_\Sigma \times \mathbf{H}_\Sigma - \frac{m_e}{e}\mathbf{j}_0 + \frac{\hbar}{4}\text{rot}(\Psi^*\boldsymbol{\sigma}\Psi) - \frac{1}{4\pi c^2}\frac{\partial G_{\Sigma k}}{\partial t}\nabla G_{\Sigma k} + \frac{1}{8\pi c^2}\text{rot}\left(\mathbf{G}_\Sigma \times \frac{\partial \mathbf{G}_\Sigma}{\partial t}\right) \qquad (112)$$

The components of the total momentum density of the Maxwell-Pauli field (112) are easily interpreted: $\mathbf{P}_{EH} = \frac{1}{4\pi c}\mathbf{E}_\Sigma \times \mathbf{H}_\Sigma$ is the momentum density of the electromagnetic field; $\mathbf{P}_e = -\frac{m_e}{e}\mathbf{j}_0 + \frac{\hbar}{4}\text{rot}(\Psi^*\boldsymbol{\sigma}\Psi)$ is the momentum density of the electron field (Pauli field), which in turn has two components: convective one $-\frac{m_e}{e}\mathbf{j}_0$ and spin one $\frac{\hbar}{4}\text{rot}(\Psi^*\boldsymbol{\sigma}\Psi)$; $\mathbf{P}_G = -\frac{1}{4\pi c^2}\frac{\partial G_{\Sigma k}}{\partial t}\nabla G_{\Sigma k} + \frac{1}{8\pi c^2}\text{rot}\left(\mathbf{G}_\Sigma \times \frac{\partial \mathbf{G}_\Sigma}{\partial t}\right)$ is the momentum density of the **G**-field, which also has two components: convective one $-\frac{1}{4\pi c^2}\frac{\partial G_{\Sigma k}}{\partial t}\nabla G_{\Sigma k}$ and spin one $\frac{1}{8\pi c^2}\text{rot}\left(\mathbf{G}_\Sigma \times \frac{\partial \mathbf{G}_\Sigma}{\partial t}\right)$.

It is easy to see that the total momentum density of the Maxwell-Pauli field (111), (112) also satisfies the continuity equation (58).

Indeed, taking into account that the rot operator of any vector can be represented as a tensor divergence, for example

$$\text{rot}(\Psi^*\boldsymbol{\sigma}\Psi)_i = \varepsilon_{ikm}\frac{\partial \Psi^*\sigma_m\Psi}{\partial x_k} = \frac{\partial \varepsilon_{ikm}\Psi^*\sigma_m\Psi}{\partial x_k}$$

and using equation (58), one obtains

$$\frac{\partial P_{\Sigma i}}{\partial t} + \frac{\partial \sigma_{ik}^{(\Sigma)}}{\partial x_k} = 0 \qquad (113)$$

where

$$\sigma_{ik}^{(\Sigma)} = \sigma_{ik} - \frac{\hbar}{4}\varepsilon_{ikm}\frac{\partial \Psi^*\sigma_m\Psi}{\partial t} - \frac{1}{8\pi c^2}\left(G_{\Sigma i}\frac{\partial^2 G_{\Sigma k}}{\partial t^2} - G_{\Sigma k}\frac{\partial^2 G_{\Sigma i}}{\partial t^2}\right) \qquad (114)$$

From (113) it follows that the total momentum of the Maxwell-Pauli field $\int \mathbf{P}_\Sigma dV$ is conserved.

Let us consider another derivation of the momentum conservation law (113) as applied to an atom in an external electromagnetic field.

Here, as before, the Maxwell-Pauli field is considered classical, and is described by equations (15)-(23), while the nucleus is considered a point classical electric charge, the motion of which is described within the framework of classical mechanics.

Considering an atom, we will explicitly take into account the presence in the region of space under consideration of a moving point charge, which is the nucleus of an atom. In this case, Maxwell equations (19)-(21) take the form

$$\text{rot}\mathbf{H}_\Sigma = \frac{1}{c}\frac{\partial \mathbf{E}_\Sigma}{\partial t} + \frac{4\pi}{c}\mathbf{j}_\Sigma \qquad (115)$$

$$\text{div}\mathbf{E}_\Sigma = 4\pi\rho_\Sigma \qquad (116)$$

$$\text{rot}\mathbf{E}_\Sigma = -\frac{1}{c}\frac{\partial \mathbf{H}_\Sigma}{\partial t} \tag{117}$$

$$\text{div}\mathbf{H}_\Sigma = 0 \tag{118}$$

where equations (117) and (118), as usual, follow from equations (21);

$$\rho_\Sigma = \rho + Ze\delta(\mathbf{r}-\mathbf{r}_n), \ \mathbf{j}_\Sigma = \mathbf{j} + Ze\mathbf{V}\delta(\mathbf{r}-\mathbf{r}_n) \tag{119}$$

is the charge density and electric current density in the area under consideration; $Ze$ is the charge of the nucleus, $\mathbf{r}_n$ is the radius-vector of the nucleus, $\mathbf{V} = \dot{\mathbf{r}}_n$ is the velocity of the nucleus. Thus, the charge density $\rho_\Sigma$ and electric current density $\mathbf{j}_\Sigma$ include both the components (22) and (23) associated with the electron field (Pauli field) and the components associated with the electric charge of the atomic nucleus.

Using equations (115)-(118), as usual [29], one obtains (see Appendix)

$$\frac{\partial}{\partial t}\left(\frac{1}{4\pi c}\mathbf{E}_\Sigma \times \mathbf{H}_\Sigma\right)_i + \frac{\partial}{\partial x_k}\frac{1}{4\pi}\left(-E_{\Sigma i}E_{\Sigma k} - H_{\Sigma i}H_{\Sigma k} + \frac{1}{2}\delta_{ik}(\mathbf{E}_\Sigma^2 + \mathbf{H}_\Sigma^2)\right) = -\left(\rho_\Sigma \mathbf{E}_\Sigma + \frac{1}{c}\mathbf{j}_\Sigma \times \mathbf{H}_\Sigma\right)_i \tag{120}$$

Multiply equation (16) for $G_{\Sigma k}$ by $\nabla G_{\Sigma k}$ and sum over $k$. As a result, one obtains

$$\frac{1}{c^2}\frac{\partial}{\partial t}\left(\frac{\partial G_{\Sigma k}}{\partial t}\nabla G_{\Sigma k}\right) - \frac{\partial}{\partial x_s}\left(\frac{\partial G_{\Sigma k}}{\partial x_s}\nabla G_{\Sigma k}\right) + \nabla\left(\frac{1}{2}\frac{\partial G_{\Sigma k}}{\partial x_s}\frac{\partial G_{\Sigma k}}{\partial x_s} - \frac{1}{2c^2}\frac{\partial G_{\Sigma k}}{\partial t}\frac{\partial G_{\Sigma k}}{\partial t}\right) = 4\pi e(\Psi^*\sigma_k\Psi)\nabla G_{\Sigma k} \tag{121}$$

or

$$\frac{\partial}{\partial t}\left(-\frac{1}{4\pi c^2}\frac{\partial G_{\Sigma k}}{\partial t}\frac{\partial G_{\Sigma k}}{\partial x_i}\right) + \frac{\partial}{\partial x_s}\frac{1}{4\pi}\left(\frac{\partial G_{\Sigma k}}{\partial x_s}\frac{\partial G_{\Sigma k}}{\partial x_i} - \delta_{is}\left(\frac{1}{2}\frac{\partial G_{\Sigma k}}{\partial x_l}\frac{\partial G_{\Sigma k}}{\partial x_l} - \frac{1}{2c^2}\frac{\partial G_{\Sigma k}}{\partial t}\frac{\partial G_{\Sigma k}}{\partial t}\right)\right) = -e(\Psi^*\sigma_k\Psi)\nabla G_{\Sigma k} \tag{122}$$

Using (40), one obtains (see Appendix)

$$-\frac{m_e}{e}\frac{\partial j_{0i}}{\partial t} = -\frac{\partial \Pi_{ik}}{\partial x_k} + \rho E_{\Sigma i} + \frac{1}{c}(\mathbf{j}_0 \times \mathbf{H}_\Sigma)_i + (\Psi^*\sigma_s\Psi)\frac{\partial}{\partial x_i}(-\mu_B H_{\Sigma s} + eG_{\Sigma s}) \tag{123}$$

where

$$\Pi_{ik} = \frac{\hbar^2}{2m_e}\left(\frac{\partial\Psi^*}{\partial x_k}\frac{\partial\Psi}{\partial x_i} + \frac{\partial\Psi^*}{\partial x_i}\frac{\partial\Psi}{\partial x_k}\right) - \frac{1}{c}(j_{0i}A_{\Sigma k} + j_{0k}A_{\Sigma i}) - \frac{e^2}{m_e c^2}A_{\Sigma i}A_{\Sigma k}\Psi^*\Psi - \delta_{ik}\frac{\hbar^2}{4m_e}\Delta(\Psi^*\Psi) \tag{124}$$

Differentiating (112) with respect to time, and taking into account (39), (41), (120), (119), (122), and (123), one obtains (see Appendix)

$$\frac{\partial P_{\Sigma i}}{\partial t} + \frac{\partial \sigma_{ik}^{(\Sigma)}}{\partial x_k} = -Ze\delta(\mathbf{r}-\mathbf{r}_n)\left(\mathbf{E}_\Sigma + \frac{1}{c}\mathbf{V}\times\mathbf{H}_\Sigma\right)_i \tag{125}$$

where

$$\sigma_{ik}^{(\Sigma)} = \frac{1}{4\pi}\left(-E_{\Sigma i}E_{\Sigma k} - H_{\Sigma i}H_{\Sigma k} + \frac{1}{2}\delta_{ik}(\mathbf{E}_\Sigma^2 + \mathbf{H}_\Sigma^2)\right) + \Pi_{ik} + \frac{1}{4\pi}\frac{\partial G_{\Sigma s}}{\partial x_k}\frac{\partial G_{\Sigma s}}{\partial x_i} - \delta_{ik}\frac{1}{8\pi}\left(\frac{\partial G_{\Sigma s}}{\partial x_m}\frac{\partial G_{\Sigma s}}{\partial x_m} - \frac{1}{c^2}\frac{\partial G_{\Sigma s}}{\partial t}\frac{\partial G_{\Sigma s}}{\partial t}\right) - \varepsilon_{iks}\frac{\hbar}{4}\frac{\partial(\Psi^*\sigma_s\Psi)}{\partial t} + \frac{1}{8\pi c^2}\left(\frac{\partial}{\partial t}\left(G_{\Sigma k}\frac{\partial G_{\Sigma i}}{\partial t}\right) - \frac{\partial}{\partial t}\left(G_{\Sigma i}\frac{\partial G_{\Sigma k}}{\partial t}\right)\right) + \delta_{ik}\mu_B\Psi^*\boldsymbol{\sigma}\Psi\mathbf{H}_\Sigma - \mu_B\Psi^*\sigma_i\Psi H_{\Sigma k} \tag{126}$$

Taking into account the classical equation of motion of the nucleus under the action of the Lorentz force

$$m_n \dot{\mathbf{V}} = Ze\left(\mathbf{E}_\Sigma + \frac{1}{c}\mathbf{v}_n \times \mathbf{H}_\Sigma\right) \tag{127}$$

where $m_n$ is the mass of the nucleus, we can write equation (125) as

$$\frac{\partial(P_{\Sigma i} + m_n V_i \delta(\mathbf{r}-\mathbf{r}_n))}{\partial t} + \frac{\partial\left(\sigma_{ik}^{(\Sigma)} + m_n V_i V_k \delta(\mathbf{r}-\mathbf{r}_n)\right)}{\partial x_k} = 0 \tag{128}$$

where $P_{\Sigma i} + m_n V_i \delta(\mathbf{r} - \mathbf{r}_n)$ and $\sigma_{ik}^{(\Sigma)} + m_n V_i V_k \delta(\mathbf{r} - \mathbf{r}_n)$ are the momentum density and the stress tensor of the combined system consisting of the Maxwell-Pauli field and the atomic nucleus. Equation (128) implies that the momentum $\int \mathbf{P}_\Sigma dV + m_n \mathbf{V}$ of the combined system is conserved, which was to be expected.

### 3.9. Movement of atoms and ions in an external electromagnetic field

In many experiments with moving atoms and ions, they behave like classical particles moving along classical trajectories, although their internal structure and interaction with external fields are described by the laws of quantum mechanics. For example, in the Stern and Gerlach experiment, the trajectories of atoms in an inhomogeneous magnetic field are well described using the laws of classical mechanics and classical electrodynamics, but the interaction of an atom with a magnetic field (in particular, only two possible orientations of the intrinsic magnetic moment of an atom relative to the magnetic field strength vector) is determined by the laws of quantum mechanics. Therefore, it is of interest to derive the equations of motion of atoms and ions as classical particles, using their description as quantum objects. In this regard, we can mention the well-known Ehrenfest theorem, but it is rather a demonstration of the connection between classical and quantum mechanics than a derivation of the laws of motion of atoms in an external field.

Consider the momentum density of an atom as a system consisting of a nucleus and a classical electron field (Pauli field) surrounding it:

$$\mathbf{P}_a = \mathbf{P}_e + m_n \mathbf{V}\delta(\mathbf{r} - \mathbf{r}_n) \tag{129}$$

After simple calculations (see Appendix), one obtains

$$\frac{\partial P_{ai}}{\partial t} + \frac{\partial K_{ik}}{\partial x_k} = \left(\rho_\Sigma \mathbf{E}_\Sigma + \frac{1}{c}\mathbf{j}_\Sigma \times \mathbf{H}_\Sigma\right)_i + e(\Psi^* \sigma_s \Psi)\frac{\partial G_{\Sigma s}}{\partial x_i} \tag{130}$$

where

$$K_{ik} = \Pi_{ik} + m_n V_i V_k \delta(\mathbf{r} - \mathbf{r}_n) - \varepsilon_{iks}\frac{\hbar}{4}\frac{\partial(\Psi^*\sigma_s\Psi)}{\partial t} + \delta_{ik}\mu_B(\Psi^*\boldsymbol{\sigma}\Psi)\mathbf{H}_\Sigma - \mu_B H_{\Sigma k}\Psi^*\sigma_i\Psi \tag{131}$$

Assuming that

$$\oint K_{ik} df_k = 0 \tag{132}$$

where the integral is taken over a sphere of infinite radius centered at the atomic nucleus, one obtains

$$\frac{\partial}{\partial t}\int \mathbf{P}_a dV = \int \left(\rho_\Sigma \mathbf{E}_\Sigma + \frac{1}{c}\mathbf{j}_\Sigma \times \mathbf{H}_\Sigma\right) dV + \int e(\Psi^* \sigma_s \Psi)\nabla G_{\Sigma s} dV \qquad (133)$$

In this section, the external electromagnetic field $\mathbf{E}, \mathbf{H}$ is considered to be the field created by external (relative to the atom) sources. Therefore, instead of relations (17), in this section we use the relations

$$\mathbf{H}_\Sigma = \mathbf{H} + \mathbf{H}_a, \quad \mathbf{E}_\Sigma = \mathbf{E} + \mathbf{E}_a, \quad \mathbf{G}_\Sigma = \mathbf{G} + \mathbf{G}_a \qquad (134)$$

where

$$\mathbf{H}_a = \mathbf{H}_n + \mathbf{H}_e, \quad \mathbf{E}_a = \mathbf{E}_n + \mathbf{E}_e, \quad \mathbf{G}_a = \mathbf{G}_n + \mathbf{G}_e \qquad (135)$$

$\mathbf{E}_n$ and $\mathbf{H}_n$ is the electromagnetic field created by the nucleus; $\mathbf{E}_e$ and $\mathbf{H}_e$ is the electromagnetic field created by the electrically charged Pauli field (electron wave) of the atom; $\mathbf{G}_n$ and $\mathbf{G}_e$ are the $\mathbf{G}$-fields, generated by the nucleus and the Pauli field, respectively.

Taking into account (134), one obtains

$$\frac{\partial}{\partial t}\int \mathbf{P}_a dV =$$
$$\int \rho_\Sigma \mathbf{E} dV + \frac{1}{c}\int (\mathbf{j}_\Sigma \times \mathbf{H}) dV + \int \left(\rho_\Sigma \mathbf{E}_a + \frac{1}{c}\mathbf{j}_\Sigma \times \mathbf{H}_a\right) dV + \int e(\Psi^* \sigma_s \Psi)\nabla G_s dV +$$
$$\int e(\Psi^* \sigma_s \Psi)\nabla G_{as} dV \qquad (136)$$

If the external electric and magnetic fields change weakly at distances of the order of the characteristic dimensions of an atom (ion), i.e., at the characteristic distances of changes in the electron field $\Psi$, we can approximately write

$$\mathbf{E} = \mathbf{E}_0 + ((\mathbf{r} - \mathbf{r}_n)\nabla)\mathbf{E}, \quad \mathbf{H} = \mathbf{H}_0 + ((\mathbf{r} - \mathbf{r}_n)\nabla)\mathbf{H} \qquad (137)$$

where $\mathbf{E}_0$ and $\mathbf{H}_0$ are the strengths of the electric and magnetic fields at the location of the atomic nucleus; field gradients within an atom are assumed to be constant. Similarly, we assume that the gradient of the external field $\mathbf{G}$ does not change at distances of the order of the characteristic dimensions of an atom (ion)

Then, taking into account (93), equation (136) takes the form

$$\frac{\partial}{\partial t}\int \mathbf{P}_a dV = \mathbf{E}_0 \int \rho_\Sigma dV + (\int (\mathbf{r}-\mathbf{r}_n)\rho_\Sigma dV \, \nabla)\mathbf{E} + \frac{1}{c}\int \mathbf{j}_\Sigma dV \times \mathbf{H}_0 + \frac{1}{c}\int (\mathbf{j}_\Sigma \times ((\mathbf{r}-\mathbf{r}_n)\nabla)\mathbf{H}) dV +$$
$$\int \left(\rho_\Sigma \mathbf{E}_a + \frac{1}{c}\mathbf{j}_\Sigma \times \mathbf{H}_a\right) dV + \frac{2e}{\hbar}\nabla(\mathbf{SG}) + \int e(\Psi^* \sigma_s \Psi)\nabla G_{as} dV \quad (138)$$

Taking into account the continuity equation (37), which is also valid for the total charge and current densities $\rho_\Sigma$ and $\mathbf{j}_\Sigma$, one obtains

$$\int \mathbf{j}_\Sigma dV = \frac{d}{dt}\int \mathbf{r}\rho_\Sigma dV + \oint \mathbf{r}(\mathbf{j}_\Sigma d\mathbf{f}) = \frac{d}{dt}\int (\mathbf{r}-\mathbf{r}_n)\rho_\Sigma dV + \frac{d}{dt}(\mathbf{r}_n \int \rho_\Sigma dV) + \oint (\mathbf{r}-\mathbf{r}_n)(\mathbf{j}_\Sigma d\mathbf{f}) +$$
$$\mathbf{r}_n \oint \mathbf{j}_\Sigma d\mathbf{f} = \dot{\mathbf{d}}_\Sigma + q_\Sigma \mathbf{V} + \oint (\mathbf{r}-\mathbf{r}_n)(\mathbf{j}_\Sigma d\mathbf{f}) \qquad (139)$$

where

$$\mathbf{d}_\Sigma = \int (\mathbf{r} - \mathbf{r}_n)\rho_\Sigma dV \qquad (140)$$

is the total dipole moment of an atom or ion with respect to the nucleus;

$$q_\Sigma = \int \rho_\Sigma dV \qquad (141)$$

is the total electric charge of the system consisting of the nucleus of an atom and the surrounding electron field. For atom $q_\Sigma = 0$; for ion $q_\Sigma \neq 0$. In (139), it is taken into account that $\dot{q}_\Sigma + \oint \mathbf{j}_\Sigma d\mathbf{f} = 0$.

Note that during the ionization of an atom, the current density $\mathbf{j}$ of the electron wave decreases at infinity as $|\mathbf{r} - \mathbf{r}_n|^{-2}$, so the integral $\oint \mathbf{j} d\mathbf{f}$ taken over the surface of a sphere of infinite radius centered at the nucleus has a finite value and is equal to the ionization current. In the absence of ionization, the electron wave current density $\mathbf{j}$ decreases at infinity faster than $|\mathbf{r} - \mathbf{r}_n|^{-2}$. Here and below, it is assumed that there is no ionization; i.e., charge $q_\Sigma$ remains constant: $\dot{q}_\Sigma = 0$. In this case, the last term on the right-hand side of (139) is equal to zero.

After simple transformations (see Appendix), we can write equation (138) as

$$\frac{\partial}{\partial t}\int \mathbf{P}_a dV = q_\Sigma \mathbf{E}_0 + q_\Sigma \frac{1}{c}\mathbf{V} \times \mathbf{H}_0 + (\mathbf{d}_\Sigma \nabla)\mathbf{E} + \nabla(\mathbf{MH}) + \frac{2e}{\hbar}\nabla(\mathbf{SG}) + \mathbf{F}_{add} \qquad (142)$$

where

$$\mathbf{F}_{add} = \frac{1}{c}\dot{\mathbf{d}}_\Sigma \times \mathbf{H}_0 + \mathbf{F}_D + \frac{1}{6c^2}\frac{\partial \mathbf{E}}{\partial t}\frac{d}{dt}\int \xi^2 \rho_\Sigma dV + \frac{1}{2c}(\mathbf{V}\nabla)(\mathbf{d}_\Sigma \times \mathbf{H}) + \frac{1}{2c}(\mathbf{d}_\Sigma \nabla)(\mathbf{V} \times \mathbf{H}) \qquad (143)$$

$$F_{Di} = \frac{1}{6c}\varepsilon_{ikm}\frac{\partial H_m}{\partial x_s}\dot{D}_{ks} \qquad (144)$$

$$D_{ks} = \int(3\xi_s\xi_k - \xi^2 \delta_{sk})\rho_\Sigma dV \qquad (145)$$

is the quadrupole moment of the electron field with respect to the nucleus.

The first four terms on the right-hand side of (142) are the usual classical forces acting on distributed classical electric charges and currents in a classical electromagnetic field [29]. The terms denoted as $\mathbf{F}_{add}$ (143), (144) are small classical additions, which, apparently, can play a significant role only in special cases. For example, the first term in (143) describes the interaction of an atom with an external magnetic field due to a rapid change (for example, rotation or as a result of spontaneous emission [22]) of the electric dipole moment of the atom. At the same time, a new force $\frac{2e}{\hbar}\nabla(\mathbf{SG})$ appeared in equation (142), which acts on the spin $\mathbf{S}$ of the Pauli field in an inhomogeneous external field $\mathbf{G}$.

## 4. Concluding remarks

Thus, we have shown that the Maxwell and Pauli fields can be naturally and consistently combined in the spirit of classical field theory without any quantization. Moreover, all the parameters of this fields, which are traditionally considered purely quantum properties and do not have a classical explanation or even an analogue, in this theory have a simple and clear physical meaning.

For such a combined theory to be self-consistent, we had to introduce a new classical field **G**. This raises a number of questions: (i) what is the field **G**? (ii) can the field **G** be included in the electromagnetic field as some component or is it a separate type of physical field? (iii) if the field **G** really exists, then why has it not yet been registered by physical devices and has not manifested itself in various experiments? (iv) how can one experimentally measure (register) the field **G**? (v) if the field **G** really exists and, as shown in this work, transfers energy and its own magnetic moment, then can it be used to transfer energy and information, similar to how it is done with the help of electromagnetic waves? We will try to answer these and other questions in the following papers of this series, but here we will give only a brief explanation.

As shown in this paper, the field **G** interacts very weakly with matter and practically does not interact with electric charges and currents. For this reason, it cannot be registered by modern physical devices, the principle of operation of which is based on the force interaction of fields with electric charges and currents. At the same time, according to the considered theory, the field **G** interacts with the intrinsic (spin) magnetic moment of the electron wave, but does not interact with the convective magnetic moments. This fact can be used as a basis for creating devices for recording and emitting the wave field **G**.

Note that even if the direct registration of the field **G** with the help of macroscopic instruments will be unsuccessful, this does not mean that it does not exist. This may mean that it is short-range and manifests itself only inside the atom. In this case, equation (16), which describes the field **G**, must be modified, but in such a way that the field **G** still compensates for the self-electrostatic field of the electron wave in the Pauli equation (15). For this purpose, the inhomogeneous Klein-Gordon equation is suitable

$$\frac{1}{c^2}\frac{\partial^2 \mathbf{G}_\Sigma}{\partial t^2} - \Delta \mathbf{G}_\Sigma + \varkappa^2 \mathbf{G}_\Sigma = 4\pi e(\Psi^*\boldsymbol{\sigma}\Psi) \qquad (146)$$

where $\varkappa$ is a constant that has the dimension inverse to the length.

The solution of the stationary ($\partial \mathbf{G}_\Sigma/\partial t = 0$) equation (146) for a localized electron field in an infinite empty ($\mathbf{G} = 0$) space, disappearing at infinity, has the form

$$\mathbf{G}_\Sigma = \mathbf{G}_e = e\int(\Psi^*\boldsymbol{\sigma}\Psi)\frac{\exp(-\varkappa R)}{R}dV' \qquad (147)$$

For small $\varkappa R \ll 1$, from (147) on approximately obtains

$$\mathbf{G}_\Sigma = \mathbf{G}_e = e\int\frac{(\Psi^*\boldsymbol{\sigma}\Psi)}{R}dV' - \varkappa e\int(\Psi^*\boldsymbol{\sigma}\Psi)dV' \qquad (148)$$

For this approximation to take place inside the atom, the condition

$$\varkappa a_B \ll 1 \qquad (149)$$

where $a_B$ is the Bohr radius.

Under condition (149), the field **G** described by equation (146), like the field (29), compensates for the own electrostatic field of the electron wave in the Pauli equation (15) and, thereby, provides the correct spontaneous emission spectrum of the hydrogen atom in the framework of the self-consistent Maxwell-Pauli theory (the second term on the right-hand side of equation (148) leads to an equidistant shift of all eigenfrequencies of the Schrödinger equation, and, therefore, does not change the spontaneous emission spectrum compared to equation (16)). At the same time, such a **G**-field is short-range one, which allows explaining the fact that this field was not previously recorded experimentally. Moreover, the short-range **G**-field allows explaining why the electron field inside the atom does not "feel" its own electrostatic field, but "feels" external electrostatic fields, for example, created by electron waves related to other atoms or ions. Thus, the short-range **G**-field allows answering the question raised at the beginning of this paper: how does an electron wave "distinguish" its own electromagnetic field from the electromagnetic field created by other electron waves (free or belonging to other atoms or ions).

From a mathematical point of view, adding a term $\varkappa^2 \mathbf{G}_\Sigma$ to equation (146) means adding a term $-\frac{1}{8\pi}\varkappa^2 \mathbf{G}_\Sigma^2$ to Lagrangian (48). This will result in adding the term $\frac{1}{8\pi}\varkappa^2 \mathbf{G}_\Sigma^2$ to the energy density (63) and (67) and the term $-\delta_{ik}\frac{1}{8\pi}\varkappa^2 \mathbf{G}_\Sigma^2$ to the stress tensor (62). The remaining components of the stress-energy tensor of the Maxwell-Pauli field will not change, just as the main fundamental results of this work will not change.

Such a version of the theory under consideration will be considered in detail in the following papers of this series.

At the same time, it should be noted that the theory considered in this paper does not contain any uncertain parameters, while the theory based on equation (146) contains a new parameter $\varkappa$, which cannot be determined within the framework of this theory. It should either be considered a new physical parameter, or, more preferably, it should be determined in terms of other parameters of the theory (for example, $a_B$ and $\alpha = \frac{e^2}{\hbar c}$) based on some additional considerations.

We note one more, as it seems to us, more general form of writing equations (15) and (146), which can be obtained by redefining the field $\mathbf{G}_\Sigma$ using the factor $-\frac{\hbar}{2m_e c}$:

$$i\hbar \frac{\partial \Psi}{\partial t} = \left[\frac{1}{2m_e}\left(\frac{\hbar}{i}\nabla + \frac{e}{c}\mathbf{A}_\Sigma\right)^2 - e\varphi_\Sigma + e\frac{2m_e c}{\hbar}\boldsymbol{\sigma}\mathbf{G}_\Sigma + \frac{e\hbar}{2m_e c}\boldsymbol{\sigma}\mathbf{H}_\Sigma\right]\Psi \qquad (150)$$

$$\frac{1}{c^2}\frac{\partial^2 \mathbf{G}_\Sigma}{\partial t^2} - \Delta \mathbf{G}_\Sigma + \varkappa^2 \mathbf{G}_\Sigma = 4\pi \mathfrak{m} \qquad (151)$$

where $\mathfrak{m}$ is the intrinsic magnetic moment density, which for an electron wave is determined by relation (41). At the same time, it is logical to assume that the magnetic moment of the nucleus also contributes to the field **G** in equation (150). In this case, the field **G** is determined by equation (151), where

$$\mathfrak{m} = \mathfrak{m}_e + \mathfrak{m}_N \tag{152}$$

$$\mathfrak{m}_e = -\mu_B \Psi^* \boldsymbol{\sigma} \Psi \tag{153}$$

is the density of the intrinsic magnetic moment of the electron field; $\mathfrak{m}_N$ is the density of the intrinsic magnetic moment of the nucleus. Taking into account the small size of the nucleus compared to the characteristic size $a_B$ of the atom, we can approximately assume that

$$\mathfrak{m}_N = \mu_N \mathbf{v}_N \delta(\mathbf{r})$$

where $\mu_N$ is the intrinsic magnetic moment of the nucleus; $\mathbf{v}_N$ is a unit vector indicating the direction of the nucleus spin. In particular, for the hydrogen atom.

$$\mathfrak{m} = \mathfrak{m}_e + \mathfrak{m}_p \tag{154}$$

where

$$\mathfrak{m}_p = \mu_p \mathbf{v}_p \delta(\mathbf{r}) \tag{155}$$

is the density of the intrinsic magnetic moment of the proton; $\mu_p = g_p \frac{e\hbar}{2m_p c}$ is the magnetic moment of the proton; $m_p$ is the proton mass; $g_p = 2.79$; $\mathbf{v}_p$ is a unit vector indicating the direction of the proton spin: $\mathbf{S}_p = \mathbf{v}_p \frac{\hbar}{2}$. Taking into account that $\mu_p \ll \mu_B$, the contribution of the proton intrinsic magnetic moment to the field **G** of the hydrogen atom is much smaller than the contribution of the electron intrinsic magnetic moment. Note that $\mu_p/\mu_B = 0.00152$, i.e., it is of the same order as the anomalous magnetic moment of the electron. Therefore, taking into account $\mathfrak{m}_p$ in equation (151) results in small corrections in the solution of equation (150) of the same order as the contribution of the anomalous magnetic moment of the electron.

**Acknowledgments**

**APPENDIX**

**A. Energy density and energy flux density**

Using (52) and (53), and taking into account (21), one obtains

$$W = \frac{1}{8\pi}\left(\mathbf{E}_\Sigma^2 + \mathbf{H}_\Sigma^2\right) + m_e c^2 \Phi^* \Phi + \frac{1}{2m_e}\left(-\frac{\hbar}{i}\nabla\Phi^* + \frac{e}{c}\mathbf{A}_\Sigma \Phi^*\right)\left(\frac{\hbar}{i}\nabla\Phi + \frac{e}{c}\mathbf{A}_\Sigma \Phi\right) - e\Phi^*\Phi\varphi_\Sigma - e\Phi^*\boldsymbol{\sigma}\Phi\mathbf{G} + \frac{e\hbar}{2m_e c}\mathbf{A}_\Sigma \mathrm{rot}(\Phi^*\boldsymbol{\sigma}\Phi) + \frac{1}{8\pi}\left(\frac{1}{c^2}\frac{\partial \mathbf{G}}{\partial t}\frac{\partial \mathbf{G}}{\partial t} + \frac{\partial \mathbf{G}}{\partial x_k}\frac{\partial \mathbf{G}}{\partial x_k}\right) + \frac{1}{4\pi}\mathbf{E}_\Sigma \nabla\varphi_\Sigma$$

Using equation (20), (22) and the obvious vector equality

$$\mathbf{A}_\Sigma \operatorname{rot}(\Phi^* \boldsymbol{\sigma} \Phi) = -\operatorname{div}(\mathbf{A}_\Sigma \times \Phi^* \boldsymbol{\sigma} \Phi) + \Phi^* \boldsymbol{\sigma} \Phi \mathbf{H}_\Sigma$$

and passing from the spinor $\Phi$ to the spinor $\Psi$ by formula (51), one obtains

$$W = \frac{1}{8\pi}\left(\mathbf{E}_\Sigma^2 + \mathbf{H}_\Sigma^2\right) - \frac{m_e c^2}{e}\rho + \frac{1}{2m_e}\left(-\frac{\hbar}{i}\nabla\Psi^* + \frac{e}{c}\mathbf{A}_\Sigma \Psi^*\right)\left(\frac{\hbar}{i}\nabla\Psi + \frac{e}{c}\mathbf{A}_\Sigma \Psi\right) + \Psi^* \boldsymbol{\sigma} \Psi(\mu_B \mathbf{H}_\Sigma -$$

$$eG) + \frac{1}{8\pi}\left(\frac{1}{c^2}\frac{\partial \mathbf{G}}{\partial t}\frac{\partial \mathbf{G}}{\partial t} + \frac{\partial \mathbf{G}}{\partial x_k}\frac{\partial \mathbf{G}}{\partial x_k}\right) + \operatorname{div}\left(\frac{1}{4\pi}\varphi_\Sigma \mathbf{E}_\Sigma - \mu_B \mathbf{A}_\Sigma \times \Psi^* \boldsymbol{\sigma} \Psi\right) \quad (A1)$$

Using (52) and (54), and taking into account that

$$H_{\Sigma k} = \varepsilon_{kls}\frac{\partial A_{\Sigma s}}{\partial x_l} \quad (A2)$$

and

$$\mathbf{A}_\Sigma \operatorname{rot}(\Phi^* \boldsymbol{\sigma} \Phi) = A_{\Sigma s}\varepsilon_{slm}\frac{\partial \Phi^* \sigma_m \Phi}{\partial x_l} = A_{\Sigma s}\varepsilon_{slm}\frac{\partial \Phi^*}{\partial x_l}\sigma_m \Phi + A_{\Sigma s}\varepsilon_{slm}\Phi^* \sigma_m \frac{\partial \Phi}{\partial x_l} \quad (A3)$$

where $\varepsilon_{slm}$ is the unit antisymmetric tensor, one obtains

$$\mathbf{J} = \frac{c}{4\pi}\mathbf{E}_\Sigma \times \mathbf{H}_\Sigma - \frac{1}{4\pi}\frac{\partial \varphi_\Sigma}{\partial t}\mathbf{E}_\Sigma + \frac{c}{4\pi}\nabla\varphi_\Sigma \times \mathbf{H}_\Sigma - \frac{\hbar}{2m_e i}\left(-\frac{\hbar}{i}\nabla\Phi^* + \frac{e}{c}\mathbf{A}_\Sigma \Phi^*\right)\frac{\partial \Phi}{\partial t} + \frac{\hbar}{2m_e i}\frac{\partial \Phi^*}{\partial t}\left(\frac{\hbar}{i}\nabla\Phi + \frac{e}{c}\mathbf{A}_\Sigma \Phi\right) + \frac{e\hbar}{2m_e c}\mathbf{A}_\Sigma \times \frac{\partial \Phi^* \boldsymbol{\sigma} \Phi}{\partial t} - \frac{1}{8\pi}\frac{\partial G_k}{\partial t}\nabla G_k$$

Let us pass from the spinor $\Phi$ to the spinor $\Psi$ using (51) and taking into account (23). As a result, one obtains

$$\mathbf{J} = \frac{c}{4\pi}\mathbf{E}_\Sigma \times \mathbf{H}_\Sigma - \frac{1}{4\pi}\frac{\partial \varphi_\Sigma}{\partial t}\mathbf{E}_\Sigma + \frac{c}{4\pi}\nabla\varphi_\Sigma \times \mathbf{H}_\Sigma - \frac{m_e c^2}{e}\mathbf{j} - \frac{\hbar c^2}{2}\operatorname{rot}(\Psi^* \boldsymbol{\sigma} \Psi) - \frac{\hbar}{2m_e i}\left(-\frac{\hbar}{i}\nabla\Psi^* + \frac{e}{c}\mathbf{A}_\Sigma \Psi^*\right)\frac{\partial \Psi}{\partial t} + \frac{\hbar}{2m_e i}\frac{\partial \Psi^*}{\partial t}\left(\frac{\hbar}{i}\nabla\Psi + \frac{e}{c}\mathbf{A}_\Sigma \Psi\right) + \frac{e\hbar}{2m_e c}\mathbf{A}_\Sigma \times \frac{\partial \Psi^* \boldsymbol{\sigma} \Psi}{\partial t} - \frac{1}{4\pi}\frac{\partial G_k}{\partial t}\nabla G_k \quad (A4)$$

Equation (57) will not change if we simultaneously make the replacement

$$W \to W - \operatorname{div}\mathbf{a}; \quad \mathbf{J} \to \mathbf{J} + \frac{\partial \mathbf{a}}{\partial t} + \operatorname{rot}\mathbf{b} \quad (A5)$$

where $\mathbf{a}$ and $\mathbf{b}$ are arbitrary vectors.

This allows redefining the parameters $W$ and $\mathbf{J}$. Choosing

$$\mathbf{a} = \frac{1}{4\pi}\varphi_\Sigma \mathbf{E}_\Sigma - \mu_B \mathbf{A}_\Sigma \times \Psi^* \boldsymbol{\sigma} \Psi; \quad \mathbf{b} = \frac{\hbar c^2}{2}\Psi^* \boldsymbol{\sigma} \Psi \quad (A6)$$

and using equation (19) we reduce the parameters (A1) and (A4) to the form

$$W = -\frac{m_e c^2}{e}\rho + W' \quad (A7)$$

$$\mathbf{J} = -\frac{m_e c^2}{e}\mathbf{j} + \mathbf{J}' \quad (A8)$$

where the components $-\frac{m_e c^2}{e}\rho$ and $-\frac{m_e c^2}{e}\mathbf{j}$ describe the rest energy density and rest energy flux density of the electron field, while

$$W' = \frac{1}{8\pi}\left(\mathbf{E}_\Sigma^2 + \mathbf{H}_\Sigma^2\right) + \frac{1}{2m_e}\left(-\frac{\hbar}{i}\nabla\Psi^* + \frac{e}{c}\mathbf{A}_\Sigma \Psi^*\right)\left(\frac{\hbar}{i}\nabla\Psi + \frac{e}{c}\mathbf{A}_\Sigma \Psi\right) - \Psi^* \boldsymbol{\sigma} \Psi(-\mu_B \mathbf{H}_\Sigma + eG) +$$

$$\frac{1}{8\pi}\left(\frac{1}{c^2}\frac{\partial \mathbf{G}}{\partial t}\frac{\partial \mathbf{G}}{\partial t} + \frac{\partial \mathbf{G}}{\partial x_k}\frac{\partial \mathbf{G}}{\partial x_k}\right) \quad (A9)$$

$$\mathbf{J}' = \frac{c}{4\pi}\mathbf{E}_\Sigma \times \mathbf{H}_\Sigma - \frac{\hbar}{2m_e i}\left(-\frac{\hbar}{i}\nabla\Psi^* + \frac{e}{c}\mathbf{A}_\Sigma\Psi^*\right)\frac{\partial\Psi}{\partial t} + \frac{\hbar}{2m_e i}\frac{\partial\Psi^*}{\partial t}\left(\frac{\hbar}{i}\nabla\Psi + \frac{e}{c}\mathbf{A}_\Sigma\Psi\right) - \mu_B\frac{\partial\mathbf{A}_\Sigma}{\partial t}\times\Psi^*\boldsymbol{\sigma}\Psi -$$

$$-\frac{1}{4\pi}\frac{\partial G_k}{\partial t}\nabla G_k - \varphi_\Sigma \mathbf{j} + \frac{c}{4\pi}\text{rot}(\varphi_\Sigma \mathbf{H}_\Sigma) \tag{A10}$$

are the non-relativistic components of the energy density and energy flux density.

Taking into account Eq. (37), it can be seen that the nonrelativistic components of the energy density and energy flux density (A9) and (A10) also satisfy the continuity equation (57), i.e., the energy conservation law.

### B. Momentum density and stress tensor

Using (52) and (55), and taking into account (21), one obtains

$$\mathbf{P} = \frac{i\hbar}{2}[(\nabla\Phi^*)\Phi - \Phi^*\nabla\Phi] + \frac{1}{4\pi c}E_{\Sigma k}\nabla A_{\Sigma k} - \frac{1}{4\pi c^2}\frac{\partial G_k}{\partial t}\nabla G_k$$

Taking into account the vector identity

$$E_{\Sigma k}\nabla A_{\Sigma k} = \mathbf{E}_\Sigma \times \text{rot}\mathbf{A}_\Sigma + (\mathbf{E}_\Sigma \nabla)\mathbf{A}_\Sigma = \mathbf{E}_\Sigma \times \mathbf{H}_\Sigma + \frac{\partial E_{\Sigma k}\mathbf{A}_\Sigma}{\partial x_k} - \mathbf{A}_\Sigma \text{div}\mathbf{E}_\Sigma,$$

the Maxwell equation (20) and relations (22), (23) and (51), one obtains

$$\mathbf{P} = \frac{1}{4\pi c}\mathbf{E}_\Sigma \times \mathbf{H}_\Sigma - \frac{m_e}{e}\mathbf{j}_0 - \frac{1}{4\pi c^2}\frac{\partial G_k}{\partial t}\nabla G_k + \frac{1}{4\pi c}\frac{\partial E_{\Sigma k}\mathbf{A}_\Sigma}{\partial x_k} \tag{B1}$$

Using (52) and (56), and taking into account (21), (A2) and (A3), one obtains

$$\sigma_{ik} = -\left(-\frac{\hbar}{2m_e i}\left(-\frac{\hbar}{i}\frac{\partial\Phi^*}{\partial x_k} + \frac{e}{c}A_{\Sigma k}\Phi^*\right)\frac{\partial\Phi}{\partial x_i} + \frac{\hbar}{2m_e i}\frac{\partial\Phi^*}{\partial x_i}\left(\frac{\hbar}{i}\frac{\partial\Phi}{\partial x_k} + \frac{e}{c}A_{\Sigma k}\Phi\right) - \frac{e\hbar}{2m_e c}A_{\Sigma s}\varepsilon_{skm}\frac{\partial\Phi^*\sigma_m\Phi}{\partial x_i} - \right.$$

$$-\frac{1}{4\pi}\left(E_{\Sigma k}\frac{\partial\varphi_\Sigma}{\partial x_i} + H_{\Sigma m}\varepsilon_{mks}\frac{\partial A_{\Sigma s}}{\partial x_i}\right) - \frac{1}{4\pi}\frac{\partial\mathbf{G}}{\partial x_i}\frac{\partial\mathbf{G}}{\partial x_k}\right) + \delta_{ik}\left(-m_e c^2\Phi^*\Phi + \frac{i\hbar}{2}\left(\Phi^*\frac{\partial\Phi}{\partial t} - \frac{\partial\Phi^*}{\partial t}\Phi\right) + \right.$$

$$\frac{1}{2m_e}\left(\frac{\hbar}{i}\nabla\Phi^* - \frac{e}{c}\mathbf{A}_\Sigma\Phi^*\right)\left(\frac{\hbar}{i}\nabla\Phi + \frac{e}{c}\mathbf{A}_\Sigma\Phi\right) + e\Phi^*\Phi\varphi_\Sigma + e\Phi^*\boldsymbol{\sigma}\Phi\mathbf{G} - \frac{e\hbar}{2m_e c}\mathbf{A}_\Sigma \text{rot}(\Phi^*\boldsymbol{\sigma}\Phi) +$$

$$\left.\frac{1}{8\pi}(\mathbf{E}_\Sigma^2 - \mathbf{H}_\Sigma^2) + \frac{1}{8\pi c^2}\frac{\partial\mathbf{G}}{\partial t}\frac{\partial\mathbf{G}}{\partial t} - \frac{1}{8\pi}\frac{\partial\mathbf{G}}{\partial x_j}\frac{\partial\mathbf{G}}{\partial x_j}\right) \tag{B2}$$

Using (19), (21), and (23), one writes

$$E_{\Sigma k}\frac{\partial\varphi_\Sigma}{\partial x_i} = -E_{\Sigma i}E_{\Sigma k} - \frac{1}{c}\frac{\partial A_{\Sigma i}}{\partial t}E_{\Sigma k}$$

$$\frac{1}{c}\frac{\partial E_{\Sigma k}}{\partial t} = (\text{rot}\mathbf{H}_\Sigma)_k - \frac{4\pi}{c}j_k = \varepsilon_{kls}\frac{\partial H_{\Sigma s}}{\partial x_l} - \frac{4\pi}{c}j_k$$

$$\frac{1}{c}A_{\Sigma i}j_k = \frac{e\hbar}{i2m_e c}A_{\Sigma i}\left(\frac{\partial\Phi^*}{\partial x_k}\Phi - \Phi^*\frac{\partial\Phi}{\partial x_k}\right) - \frac{e^2}{m_e c^2}A_{\Sigma i}A_{\Sigma k}\Phi^*\Phi - \frac{e\hbar}{2m_e c}A_{\Sigma i}\varepsilon_{klm}\frac{\partial\Phi^*\sigma_m\Phi}{\partial x_l}$$

$$\mathbf{H}_\Sigma^2 = H_{\Sigma m}\varepsilon_{mls}\frac{\partial A_{\Sigma s}}{\partial x_l}$$

$$-\frac{\hbar}{2m_e i}\left(-\frac{\hbar}{i}\frac{\partial \Phi^*}{\partial x_k}+\frac{e}{c}A_{\Sigma k}\Phi^*\right)\frac{\partial \Phi}{\partial x_i}+\frac{\hbar}{2m_e i}\frac{\partial \Phi^*}{\partial x_i}\left(\frac{\hbar}{i}\frac{\partial \Phi}{\partial x_k}+\frac{e}{c}A_{\Sigma k}\Phi\right)-\frac{e\hbar}{2m_e c}A_{\Sigma s}\varepsilon_{skm}\frac{\partial \Phi^*\sigma_m \Phi}{\partial x_i}$$

$$-\frac{1}{4\pi}\left(E_{\Sigma k}\frac{\partial \varphi_\Sigma}{\partial x_i}+H_{\Sigma m}\varepsilon_{mks}\frac{\partial A_{\Sigma s}}{\partial x_i}\right)-\frac{1}{4\pi}\frac{\partial \mathbf{G}}{\partial x_i}\frac{\partial \mathbf{G}}{\partial x_k}$$

$$=\frac{1}{4\pi}E_{\Sigma i}E_{\Sigma k}-\frac{\hbar^2}{2m_e}\left(\frac{\partial \Phi^*}{\partial x_k}\frac{\partial \Phi}{\partial x_i}+\frac{\partial \Phi^*}{\partial x_i}\frac{\partial \Phi}{\partial x_k}\right)+\frac{e\hbar}{i2m_e c}A_{\Sigma i}\left(\frac{\partial \Phi^*}{\partial x_k}\Phi-\Phi^*\frac{\partial \Phi}{\partial x_k}\right)$$

$$+\frac{\hbar e}{i2m_e c}A_{\Sigma k}\left(\frac{\partial \Phi^*}{\partial x_i}\Phi-\Phi^*\frac{\partial \Phi}{\partial x_i}\right)-\frac{e^2}{m_e c^2}A_{\Sigma i}A_{\Sigma k}\Phi^*\Phi-\frac{1}{4\pi}\frac{\partial \mathbf{G}}{\partial x_i}\frac{\partial \mathbf{G}}{\partial x_k}$$

$$-\frac{e\hbar}{2m_e c}(\delta_{il}\varepsilon_{skm}+\delta_{is}\varepsilon_{klm})A_{\Sigma s}\frac{\partial \Phi^*\sigma_m \Phi}{\partial x_l}$$

$$+\frac{1}{4\pi}(\delta_{ik}\varepsilon_{mls}+\delta_{is}\varepsilon_{klm}-\delta_{il}\varepsilon_{mks})H_{\Sigma m}\frac{\partial A_{\Sigma s}}{\partial x_l}+\frac{1}{4\pi c}\frac{\partial A_{\Sigma i}E_{\Sigma k}}{\partial t}-\frac{1}{4\pi}\frac{\partial \varepsilon_{kls}A_{\Sigma i}H_{\Sigma s}}{\partial x_l}$$

$$-\frac{1}{4\pi}\delta_{ik}\mathbf{H}_\Sigma^2$$

Taking into account the easily verifiable identity

$$\delta_{il}\varepsilon_{mks}+\delta_{im}\varepsilon_{kls}=\delta_{is}\varepsilon_{klm}+\delta_{ik}\varepsilon_{mls}$$

one obtains

$$-\frac{\hbar}{2m_e i}\left(-\frac{\hbar}{i}\frac{\partial \Phi^*}{\partial x_k}+\frac{e}{c}A_{\Sigma k}\Phi^*\right)\frac{\partial \Phi}{\partial x_i}+\frac{\hbar}{2m_e i}\frac{\partial \Phi^*}{\partial x_i}\left(\frac{\hbar}{i}\frac{\partial \Phi}{\partial x_k}+\frac{e}{c}A_{\Sigma k}\Phi\right)-\frac{e\hbar}{2m_e c}A_{\Sigma s}\varepsilon_{skm}\frac{\partial \Phi^*\sigma_m \Phi}{\partial x_i}-$$

$$\frac{1}{4\pi}\left(E_{\Sigma k}\frac{\partial \varphi_\Sigma}{\partial x_i}+H_{\Sigma m}\varepsilon_{mks}\frac{\partial A_{\Sigma s}}{\partial x_i}\right)-\frac{1}{4\pi}\frac{\partial \mathbf{G}}{\partial x_i}\frac{\partial \mathbf{G}}{\partial x_k}=\frac{1}{4\pi}E_{\Sigma i}E_{\Sigma k}+\frac{1}{4\pi}H_{\Sigma i}H_{\Sigma k}-\frac{\hbar^2}{2m_e}\left(\frac{\partial \Phi^*}{\partial x_k}\frac{\partial \Phi}{\partial x_i}+\frac{\partial \Phi^*}{\partial x_i}\frac{\partial \Phi}{\partial x_k}\right)+$$

$$\frac{e\hbar}{i2m_e c}A_{\Sigma i}\left(\frac{\partial \Phi^*}{\partial x_k}\Phi-\Phi^*\frac{\partial \Phi}{\partial x_k}\right)+\frac{\hbar e}{i2m_e c}A_{\Sigma k}\left(\frac{\partial \Phi^*}{\partial x_i}\Phi-\Phi^*\frac{\partial \Phi}{\partial x_i}\right)-\frac{e^2}{m_e c^2}A_{\Sigma i}A_{\Sigma k}\Phi^*\Phi-\frac{1}{4\pi}\frac{\partial \mathbf{G}}{\partial x_i}\frac{\partial \mathbf{G}}{\partial x_k}-$$

$$\frac{e\hbar}{2m_e c}(\delta_{il}\varepsilon_{skm}+\delta_{is}\varepsilon_{klm})A_{\Sigma s}\frac{\partial \Phi^*\sigma_m \Phi}{\partial x_l}+\frac{1}{4\pi c}\frac{\partial A_{\Sigma i}E_{\Sigma k}}{\partial t}-\frac{1}{4\pi}\frac{\partial \varepsilon_{kls}A_{\Sigma i}H_{\Sigma s}}{\partial x_l}-\frac{1}{4\pi}\delta_{ik}\mathbf{H}_\Sigma^2 \quad (B3)$$

Substituting (B3) into (B2) and going from $\Phi$ to $\Psi$, one obtains

$$\sigma_{ik}=-\frac{1}{4\pi}E_{\Sigma i}E_{\Sigma k}-\frac{1}{4\pi}H_{\Sigma i}H_{\Sigma k}+\frac{\hbar^2}{2m_e}\left(\frac{\partial \Psi^*}{\partial x_k}\frac{\partial \Psi}{\partial x_i}+\frac{\partial \Psi^*}{\partial x_i}\frac{\partial \Psi}{\partial x_k}\right)-\frac{e\hbar}{i2m_e c}A_{\Sigma i}\left(\frac{\partial \Psi^*}{\partial x_k}\Psi-\Psi^*\frac{\partial \Psi}{\partial x_k}\right)-$$

$$\frac{\hbar e}{i2m_e c}A_{\Sigma k}\left(\frac{\partial \Psi^*}{\partial x_i}\Psi-\Psi^*\frac{\partial \Psi}{\partial x_i}\right)+\frac{e^2}{m_e c^2}A_{\Sigma i}A_{\Sigma k}\Psi^*\Psi+\frac{1}{4\pi}\frac{\partial \mathbf{G}}{\partial x_i}\frac{\partial \mathbf{G}}{\partial x_k}+$$

$$\frac{e\hbar}{2m_e c}(\delta_{il}\varepsilon_{skm}+\delta_{is}\varepsilon_{klm})A_{\Sigma s}\frac{\partial \Psi^*\sigma_m \Psi}{\partial x_l}+\delta_{ik}\left(\frac{1}{8\pi}(\mathbf{E}_\Sigma^2+\mathbf{H}_\Sigma^2)+\frac{i\hbar}{2}\left(\Psi^*\frac{\partial \Psi}{\partial t}-\frac{\partial \Psi^*}{\partial t}\Psi\right)+\frac{1}{2m_e}\left(\frac{\hbar}{i}\nabla\Psi^*-\right.\right.$$

$$\left.\frac{e}{c}\mathbf{A}_\Sigma\Psi^*\right)\left(\frac{\hbar}{i}\nabla\Psi+\frac{e}{c}\mathbf{A}_\Sigma\Psi\right)+e\Psi^*\Psi\varphi_\Sigma+e\Psi^*\boldsymbol{\sigma}\Psi\mathbf{G}-\frac{e\hbar}{2m_e c}\mathbf{A}_\Sigma\text{rot}(\Psi^*\boldsymbol{\sigma}\Psi)+\frac{1}{8\pi c^2}\frac{\partial \mathbf{G}}{\partial t}\frac{\partial \mathbf{G}}{\partial t}-$$

$$\left.\frac{1}{8\pi}\frac{\partial \mathbf{G}}{\partial x_j}\frac{\partial \mathbf{G}}{\partial x_j}\right)-\frac{1}{4\pi c}\frac{\partial A_{\Sigma i}E_{\Sigma k}}{\partial t}+\frac{1}{4\pi}\frac{\partial \varepsilon_{kls}A_{\Sigma i}H_{\Sigma s}}{\partial x_l} \quad (B4)$$

Obviously, the continuity equation (58) will not change if we simultaneously change

$$P_i \to P_i+\frac{\partial u_{ik}}{\partial x_k}; \ \sigma_{ik}\to \sigma_{ik}-\frac{\partial u_{ik}}{\partial t}+\frac{\partial q_{iks}}{\partial x_s} \quad (B5)$$

where $u_{ik}$ and $q_{iks}=-q_{isk}$ are arbitrary functions.

This allows redefining the parameters **P** and $\sigma_{ik}$.

Choosing

$$u_{ik} = -\frac{1}{4\pi c}A_{\Sigma i}E_{\Sigma k}; \quad q_{iks} = -\frac{1}{4\pi}\varepsilon_{ksm}A_{\Sigma i}H_{\Sigma m} \tag{B6}$$

we reduce the parameters (B1) and (B4) to the form

$$\mathbf{P} = \frac{1}{4\pi c}\mathbf{E}_\Sigma \times \mathbf{H}_\Sigma - \frac{m_e}{e}\mathbf{j}_0 - \frac{1}{4\pi c^2}\frac{\partial G_k}{\partial t}\nabla G_k \tag{B7}$$

$$\sigma_{ik} = -\frac{1}{4\pi}E_{\Sigma i}E_{\Sigma k} - \frac{1}{4\pi}H_{\Sigma i}H_{\Sigma k} + \frac{\hbar^2}{2m_e}\left(\frac{\partial \Psi^*}{\partial x_k}\frac{\partial \Psi}{\partial x_i} + \frac{\partial \Psi^*}{\partial x_i}\frac{\partial \Psi}{\partial x_k}\right) - \frac{e\hbar}{i2m_e c}A_{\Sigma i}\left(\frac{\partial \Psi^*}{\partial x_k}\Psi - \Psi^*\frac{\partial \Psi}{\partial x_k}\right) -$$

$$\frac{\hbar e}{i2m_e c}A_{\Sigma k}\left(\frac{\partial \Psi^*}{\partial x_i}\Psi - \Psi^*\frac{\partial \Psi}{\partial x_i}\right) + \frac{e^2}{m_e c^2}A_{\Sigma i}A_{\Sigma k}\Psi^*\Psi + \frac{1}{4\pi}\frac{\partial \mathbf{G}}{\partial x_i}\frac{\partial \mathbf{G}}{\partial x_k} + \delta_{ik}\left(\frac{1}{8\pi}(\mathbf{E}_\Sigma^2 + \mathbf{H}_\Sigma^2) + \right.$$

$$\frac{i\hbar}{2}\left(\Psi^*\frac{\partial \Psi}{\partial t} - \frac{\partial \Psi^*}{\partial t}\Psi\right) + \frac{1}{2m_e}\left(\frac{\hbar}{i}\nabla\Psi^* - \frac{e}{c}\mathbf{A}_\Sigma\Psi^*\right)\left(\frac{\hbar}{i}\nabla\Psi + \frac{e}{c}\mathbf{A}_\Sigma\Psi\right) + e\Psi^*\Psi\varphi_\Sigma + e\Psi^*\boldsymbol{\sigma}\Psi\mathbf{G} -$$

$$\frac{e\hbar}{2m_e c}\mathbf{A}_\Sigma\mathrm{rot}(\Psi^*\boldsymbol{\sigma}\Psi) + \frac{1}{8\pi c^2}\frac{\partial \mathbf{G}}{\partial t}\frac{\partial \mathbf{G}}{\partial t} - \frac{1}{8\pi}\frac{\partial \mathbf{G}}{\partial x_j}\frac{\partial \mathbf{G}}{\partial x_j}\right) + \frac{e\hbar}{2m_e c}(\delta_{il}\varepsilon_{skm} + \delta_{is}\varepsilon_{klm})A_{\Sigma s}\frac{\partial \Psi^*\sigma_m\Psi}{\partial x_l} \tag{B8}$$

### C. Convective angular momentum

Using (B8), one calculates

$$\varepsilon_{rki}(\sigma_{ik} - \sigma_{ki}) = \frac{e\hbar}{2m_e c}(\varepsilon_{rki}\delta_{il}\varepsilon_{skm} + \varepsilon_{rki}\delta_{is}\varepsilon_{klm} - \varepsilon_{rki}\delta_{kl}\varepsilon_{sim} - \varepsilon_{rki}\delta_{ks}\varepsilon_{ilm})A_{\Sigma s}\frac{\partial \Psi^*\sigma_m\Psi}{\partial x_l}$$

Taking into account the identity

$$\varepsilon_{krl}\varepsilon_{ksm} = \delta_{rs}\delta_{lm} - \delta_{rm}\delta_{ls} \tag{C1}$$

one obtains

$$\varepsilon_{rki}\delta_{il}\varepsilon_{skm} + \varepsilon_{rki}\delta_{is}\varepsilon_{klm} - \varepsilon_{rki}\delta_{kl}\varepsilon_{sim} - \varepsilon_{rki}\delta_{ks}\varepsilon_{ilm}$$

$$= \varepsilon_{rkl}\varepsilon_{skm} + \varepsilon_{rks}\varepsilon_{klm} - \varepsilon_{rli}\varepsilon_{sim} - \varepsilon_{rsi}\varepsilon_{ilm}$$

$$= \varepsilon_{krl}\varepsilon_{ksm} - \varepsilon_{krs}\varepsilon_{klm} - \varepsilon_{ilr}\varepsilon_{ism} + \varepsilon_{isr}\varepsilon_{ilm}$$

$$= \delta_{rs}\delta_{lm} - \delta_{rm}\delta_{ls} - \delta_{rl}\delta_{sm} + \delta_{rm}\delta_{ls} - \delta_{ls}\delta_{rm} + \delta_{lm}\delta_{rs} + \delta_{ls}\delta_{rm} - \delta_{sm}\delta_{lr}$$

$$= 2(\delta_{rs}\delta_{lm} - \delta_{rl}\delta_{sm})$$

Then

$$\frac{1}{2}\varepsilon_{rki}(\sigma_{ik} - \sigma_{ki}) = \frac{e\hbar}{2m_e c}(\delta_{rs}\delta_{lm} - \delta_{rl}\delta_{sm})A_{\Sigma s}\frac{\partial \Psi^*\sigma_m\Psi}{\partial x_l}$$

$$= \frac{e\hbar}{2m_e c}\left(A_{\Sigma r}\frac{\partial \Psi^*\sigma_m\Psi}{\partial x_m} - A_{\Sigma m}\frac{\partial \Psi^*\sigma_m\Psi}{\partial x_r}\right)$$

$$= \frac{e\hbar}{2m_e c}\left(\frac{\partial}{\partial x_m}(A_{\Sigma r}\Psi^*\sigma_m\Psi - \delta_{rm}A_{\Sigma s}\Psi^*\sigma_s\Psi) + (\Psi^*\sigma_m\Psi)\left(\frac{\partial A_{\Sigma m}}{\partial x_r} - \frac{\partial A_{\Sigma r}}{\partial x_m}\right)\right)$$

$$(\Psi^*\sigma_m\Psi)\left(\frac{\partial A_{\Sigma m}}{\partial x_r} - \frac{\partial A_{\Sigma r}}{\partial x_m}\right) = (\Psi^*\sigma_m\Psi)\varepsilon_{lrm}(\mathrm{rot}\mathbf{A}_\Sigma)_l = \varepsilon_{rml}(\Psi^*\sigma_m\Psi)H_{\Sigma l} = \left((\Psi^*\boldsymbol{\sigma}\Psi) \times \mathbf{H}_\Sigma\right)_r$$

As a result, one obtains

$$\frac{1}{2}\varepsilon_{ikl}(\sigma_{lk} - \sigma_{kl}) = \frac{e\hbar}{2m_e c}\left(\frac{\partial}{\partial x_k}(A_{\Sigma i}\Psi^*\sigma_k\Psi - \delta_{ik}A_{\Sigma s}\Psi^*\sigma_s\Psi) + \left((\Psi^*\boldsymbol{\sigma}\Psi) \times \mathbf{H}_\Sigma\right)_i\right) \tag{C2}$$

Taking (C2) into account, equation (83) takes the form

$$\frac{\partial l_i}{\partial t} + \frac{\partial}{\partial x_k}\left(Q_{ik} - \frac{e\hbar}{2m_e c}(A_{\Sigma i}\Psi^*\sigma_k\Psi - \delta_{ik}A_{\Sigma s}\Psi^*\sigma_s\Psi)\right) = \mu_B\big((\Psi^*\boldsymbol{\sigma}\Psi)\times \mathbf{H}_\Sigma\big)_i \qquad (C3)$$

### D. Intrinsic angular momentum of the Pauli field

Taking into account the Pauli equation (15), one obtains

$$i\hbar\frac{\partial \Psi}{\partial t} = -\frac{\hbar^2}{2m_e}\Delta\Psi + \frac{e\hbar}{i2m_e c}\nabla(\mathbf{A}_\Sigma\Psi) + \frac{e\hbar}{i2m_e c}(\mathbf{A}_\Sigma\nabla)\Psi + \frac{e^2}{2m_e c^2}\mathbf{A}_\Sigma^2\Psi - e\varphi_\Sigma\Psi - e\boldsymbol{\sigma}\mathbf{G}\Psi$$

$$+ \frac{e\hbar}{2m_e c}\boldsymbol{\sigma}\mathbf{H}_\Sigma\Psi$$

$$-i\hbar\frac{\partial \Psi^*}{\partial t} = -\frac{\hbar^2}{2m_e}\Delta\Psi^* - \frac{e\hbar}{i2m_e c}\nabla(\mathbf{A}_\Sigma\Psi^*) - \frac{e\hbar}{i2m_e c}(\mathbf{A}_\Sigma\nabla)\Psi^* + \frac{e^2}{2m_e c^2}\mathbf{A}_\Sigma^2\Psi^* - e\varphi_\Sigma\Psi^*$$

$$- e\Psi^*\boldsymbol{\sigma}\mathbf{G} + \frac{e\hbar}{2m_e c}\Psi^*\boldsymbol{\sigma}\mathbf{H}_\Sigma$$

From here, one obtains

$$i\hbar\Psi^*\boldsymbol{\sigma}\frac{\partial \Psi}{\partial t} = -\frac{\hbar^2}{2m_e}\Psi^*\boldsymbol{\sigma}\Delta\Psi + \frac{e\hbar}{i2m_e c}\Psi^*\boldsymbol{\sigma}\nabla(\mathbf{A}_\Sigma\Psi) + \frac{e\hbar}{i2m_e c}\Psi^*\boldsymbol{\sigma}(\mathbf{A}_\Sigma\nabla)\Psi + \frac{e^2}{2m_e c^2}\mathbf{A}_\Sigma^2\Psi^*\boldsymbol{\sigma}\Psi$$

$$- e\varphi_\Sigma\Psi^*\boldsymbol{\sigma}\Psi - e\Psi^*\boldsymbol{\sigma}(\boldsymbol{\sigma}\mathbf{G})\Psi + \frac{e\hbar}{2m_e c}\Psi^*\boldsymbol{\sigma}(\boldsymbol{\sigma}\mathbf{H}_\Sigma)\Psi$$

$$-i\hbar\frac{\partial \Psi^*\boldsymbol{\sigma}}{\partial t}\Psi = -\frac{\hbar^2}{2m_e}\Delta\Psi^*\boldsymbol{\sigma}\Psi - \frac{e\hbar}{i2m_e c}\nabla(\mathbf{A}_\Sigma\Psi^*)\boldsymbol{\sigma}\Psi - \frac{e\hbar}{i2m_e c}(\mathbf{A}_\Sigma\nabla)\Psi^*\boldsymbol{\sigma}\Psi + \frac{e^2}{2m_e c^2}\mathbf{A}_\Sigma^2\Psi^*\boldsymbol{\sigma}\Psi$$

$$- e\varphi_\Sigma\Psi^*\boldsymbol{\sigma}\Psi - e\Psi^*(\boldsymbol{\sigma}\mathbf{G})\boldsymbol{\sigma}\Psi + \frac{e\hbar}{2m_e c}\Psi^*(\boldsymbol{\sigma}\mathbf{H}_\Sigma)\boldsymbol{\sigma}\Psi$$

Subtracting the second one from the first one, one obtains

$$i\hbar\Psi^*\boldsymbol{\sigma}\frac{\partial \Psi}{\partial t} + i\hbar\frac{\partial \Psi^*\boldsymbol{\sigma}}{\partial t}\Psi$$

$$= -\frac{\hbar^2}{2m_e}\Psi^*\boldsymbol{\sigma}\Delta\Psi + \frac{\hbar^2}{2m_e}\Delta\Psi^*\boldsymbol{\sigma}\Psi + \frac{e\hbar}{i2m_e c}\Psi^*\boldsymbol{\sigma}\nabla(\mathbf{A}_\Sigma\Psi) + \frac{e\hbar}{i2m_e c}\nabla(\mathbf{A}_\Sigma\Psi^*)\boldsymbol{\sigma}\Psi$$

$$+ \frac{e\hbar}{i2m_e c}(\mathbf{A}_\Sigma\nabla)\Psi^*\boldsymbol{\sigma}\Psi + \frac{e\hbar}{i2m_e c}\Psi^*\boldsymbol{\sigma}(\mathbf{A}_\Sigma\nabla)\Psi - e\Psi^*\boldsymbol{\sigma}(\boldsymbol{\sigma}\mathbf{G})\Psi + e\Psi^*(\boldsymbol{\sigma}\mathbf{G})\boldsymbol{\sigma}\Psi$$

$$+ \frac{e\hbar}{2m_e c}\Psi^*\boldsymbol{\sigma}(\boldsymbol{\sigma}\mathbf{H}_\Sigma)\Psi - \frac{e\hbar}{2m_e c}\Psi^*(\boldsymbol{\sigma}\mathbf{H}_\Sigma)\boldsymbol{\sigma}\Psi$$

Taking into account the properties of Pauli matrices [28]

$$(\boldsymbol{\sigma}\mathbf{G})\boldsymbol{\sigma} = \mathbf{G} + i\boldsymbol{\sigma}\times\mathbf{G}, \qquad \boldsymbol{\sigma}(\boldsymbol{\sigma}\mathbf{G}) = \mathbf{G} + i\mathbf{G}\times\boldsymbol{\sigma}$$

one obtains

$$i\hbar \frac{\partial \Psi^* \boldsymbol{\sigma} \Psi}{\partial t} = -\frac{\hbar^2}{2m_e} \Psi^* \boldsymbol{\sigma} \Delta \Psi + \frac{\hbar^2}{2m_e} \Delta \Psi^* \boldsymbol{\sigma} \Psi + \frac{e\hbar}{i2m_e c} \Psi^* \boldsymbol{\sigma} \nabla(\mathbf{A}_\Sigma \Psi) + \frac{e\hbar}{i2m_e c} \nabla(\mathbf{A}_\Sigma \Psi^*) \boldsymbol{\sigma} \Psi$$

$$+ \frac{e\hbar}{i2m_e c}(\mathbf{A}_\Sigma \nabla) \Psi^* \boldsymbol{\sigma} \Psi + \frac{e\hbar}{i2m_e c} \Psi^* \boldsymbol{\sigma} (\mathbf{A}_\Sigma \nabla) \Psi + 2i\Psi^* \boldsymbol{\sigma} \Psi \times (-\mu_B \mathbf{H}_\Sigma + e\mathbf{G})$$

or

$$\frac{\hbar}{2} \frac{\partial \Psi^* \sigma_k \Psi}{\partial t} = -\frac{\hbar}{2m_e} \frac{\partial}{\partial x_s} \left[ \frac{\hbar}{2i} \left( \Psi^* \sigma_k \frac{\partial \Psi}{\partial x_s} - \frac{\partial \Psi^*}{\partial x_s} \sigma_k \Psi \right) + \frac{e}{c} A_{\Sigma s} \Psi^* \sigma_k \Psi \right] + K_k$$

where

$$\mathbf{K} = \Psi^* \boldsymbol{\sigma} \Psi \times (-\mu_B \mathbf{H}_\Sigma + e\mathbf{G})$$

### E. Total momentum of the Maxwell-Pauli field

Using equations (115)-(118), as usual [29], one obtains

$$\frac{1}{c} \frac{\partial \mathbf{E}_\Sigma \times \mathbf{H}_\Sigma}{\partial t} = -\mathbf{H}_\Sigma \times \text{rot}\mathbf{H}_\Sigma - \mathbf{E}_\Sigma \times \text{rot}\mathbf{E}_\Sigma - \frac{4\pi}{c} \mathbf{j}_\Sigma \times \mathbf{H}_\Sigma$$

We take into account that

$$\mathbf{H}_\Sigma \times \text{rot}\mathbf{H}_\Sigma = \frac{1}{2} \nabla \mathbf{H}_\Sigma^2 - (\mathbf{H}_\Sigma \nabla) \mathbf{H}_\Sigma$$

Then

$$\frac{1}{c} \frac{\partial (\mathbf{E}_\Sigma \times \mathbf{H}_\Sigma)_i}{\partial t} + \frac{\partial}{\partial x_k} \left( -E_{\Sigma i} E_{\Sigma k} - H_{\Sigma i} H_{\Sigma k} + \frac{1}{2} \delta_{ik} (\mathbf{E}_\Sigma^2 + \mathbf{H}_\Sigma^2) \right) = -H_{\Sigma i} \frac{\partial H_{\Sigma k}}{\partial x_k} - E_{\Sigma i} \frac{\partial E_{\Sigma k}}{\partial x_k} - \frac{4\pi}{c} (\mathbf{j} \times \mathbf{H}_\Sigma)_i$$

Taking into account (116) and (118), one obtains

$$\frac{\partial}{\partial t} \left( \frac{1}{4\pi c} \mathbf{E}_\Sigma \times \mathbf{H}_\Sigma \right)_i + \frac{\partial}{\partial x_k} \frac{1}{4\pi} \left( -E_{\Sigma i} E_{\Sigma k} - H_{\Sigma i} H_{\Sigma k} + \frac{1}{2} \delta_{ik} (\mathbf{E}_\Sigma^2 + \mathbf{H}_\Sigma^2) \right) = -\left( \rho_\Sigma \mathbf{E}_\Sigma + \frac{1}{c} \mathbf{j}_\Sigma \times \mathbf{H}_\Sigma \right)_i$$
(E1)

Using (40), one writes

$$-\frac{m_e}{e} \frac{\partial \mathbf{j}_0}{\partial t} = -\frac{\hbar}{2i} \left[ \left( \nabla \frac{\partial \Psi^*}{\partial t} \right) \Psi + (\nabla \Psi^*) \frac{\partial \Psi}{\partial t} - \frac{\partial \Psi^*}{\partial t} \nabla \Psi - \Psi^* \nabla \frac{\partial \Psi}{\partial t} \right] + \frac{e}{c} \frac{\partial \mathbf{A}_\Sigma}{\partial t} \Psi^* \Psi + \frac{e}{c} \frac{\partial \Psi^* \Psi}{\partial t} \mathbf{A}_\Sigma =$$

$$-\nabla \left( -\frac{1}{2} i\hbar \frac{\partial \Psi^*}{\partial t} \Psi + \frac{1}{2} i\hbar \Psi^* \frac{\partial \Psi}{\partial t} \right) - i\hbar \left( \frac{\partial \Psi^*}{\partial t} \right) \nabla \Psi + i\hbar (\nabla \Psi^*) \frac{\partial \Psi}{\partial t} + \frac{e}{c} \frac{\partial \mathbf{A}_\Sigma}{\partial t} \Psi^* \Psi + \frac{e}{c} \frac{\partial \Psi^* \Psi}{\partial t} \mathbf{A}_\Sigma \quad \text{(E1)}$$

Using equations (15) and (50), one obtains

$$-i\hbar \left( \frac{\partial \Psi^*}{\partial t} \right) \nabla \Psi + i\hbar (\nabla \Psi^*) \frac{\partial \Psi}{\partial t} = \left( \frac{1}{2m_e} \left( -\frac{\hbar}{i} \nabla + \frac{e}{c} \mathbf{A}_\Sigma \right)^2 \Psi^* \right) \nabla \Psi + \frac{1}{2m_e} (\nabla \Psi^*) \left( \frac{\hbar}{i} \nabla + \frac{e}{c} \mathbf{A}_\Sigma \right)^2 \Psi -$$

$$e\varphi_\Sigma \nabla(\Psi^* \Psi) - eG_s \nabla(\Psi^* \sigma_s \Psi) + \mu_B H_{\Sigma s} \nabla(\Psi^* \sigma_s \Psi) = \left( \frac{1}{2m_e} \left( -\frac{\hbar}{i} \nabla + \frac{e}{c} \mathbf{A}_\Sigma \right)^2 \Psi^* \right) \nabla \Psi +$$

$$\frac{1}{2m_e} (\nabla \Psi^*) \left( \frac{\hbar}{i} \nabla + \frac{e}{c} \mathbf{A}_\Sigma \right)^2 \Psi - \nabla(e\varphi_\Sigma \Psi^* \Psi - \mu_B H_{\Sigma s} \Psi^* \sigma_s \Psi + eG_s \Psi^* \sigma_s \Psi) + e\Psi^* \Psi \nabla \varphi_\Sigma +$$

$$\Psi^* \sigma_s \Psi \nabla(-\mu_B H_{\Sigma s} + eG_s) \quad \text{(E2)}$$

Let us transform, taking into account (40):

$$\left(\frac{1}{2m_e}\left(-\frac{\hbar}{i}\nabla + \frac{e}{c}\mathbf{A}_\Sigma\right)^2 \Psi^*\right)\nabla\Psi + \frac{1}{2m_e}(\nabla\Psi^*)\left(\frac{\hbar}{i}\nabla + \frac{e}{c}\mathbf{A}_\Sigma\right)^2 \Psi = -\frac{\hbar^2}{2m_e}\left(\frac{\partial}{\partial x_s}\frac{\partial \Psi^*}{\partial x_s}\right)\nabla\Psi -$$

$$\frac{\hbar^2}{2m_e}(\nabla\Psi^*)\frac{\partial}{\partial x_s}\frac{\partial\Psi}{\partial x_s} + \frac{e\hbar}{2m_e c i}\frac{\partial A_{\Sigma s}}{\partial x_s}\left((\nabla\Psi^*)\Psi - \Psi^*\nabla\Psi\right) - \frac{e\hbar}{m_e c i}A_{\Sigma s}\frac{\partial\Psi^*}{\partial x_s}\nabla\Psi + \frac{e\hbar}{m_e c i}(\nabla\Psi^*)A_{\Sigma s}\frac{\partial\Psi}{\partial x_s} +$$

$$\frac{e^2}{2m_e c^2}\mathbf{A}_\Sigma^2\nabla(\Psi^*\Psi) = -\frac{\hbar^2}{2m_e}\left(\frac{\partial}{\partial x_s}\frac{\partial\Psi^*}{\partial x_s}\right)\nabla\Psi - \frac{\hbar^2}{2m_e}(\nabla\Psi^*)\frac{\partial}{\partial x_s}\frac{\partial\Psi}{\partial x_s} + \frac{e\hbar}{2m_e c i}\frac{\partial A_{\Sigma s}}{\partial x_s}\left((\nabla\Psi^*)\Psi - \Psi^*\nabla\Psi\right) +$$

$$\frac{e\hbar}{m_e c i}A_{\Sigma s}\left((\nabla\Psi^*)\frac{\partial\Psi}{\partial x_s} - \frac{\partial\Psi^*}{\partial x_s}\nabla\Psi\right) + \frac{e^2}{2m_e c^2}\mathbf{A}_\Sigma^2\nabla(\Psi^*\Psi) = -\frac{\hbar^2}{2m_e}\frac{\partial}{\partial x_s}\left(\frac{\partial\Psi^*}{\partial x_s}\nabla\Psi + (\nabla\Psi^*)\frac{\partial\Psi}{\partial x_s}\right) +$$

$$\frac{\hbar^2}{2m_e}\nabla\frac{\partial\Psi^*}{\partial x_s}\frac{\partial\Psi}{\partial x_s} + \frac{1}{c}\frac{\partial}{\partial x_s}\left(A_{\Sigma s}\mathbf{j}_0 + \frac{e^2}{m_e c}A_{\Sigma s}\mathbf{A}_\Sigma\Psi^*\Psi\right) - \frac{1}{c}A_{\Sigma s}\nabla\left(j_{0s} + \frac{e^2}{m_e c}A_{\Sigma s}\Psi^*\Psi\right) +$$

$$\frac{e^2}{2m_e c^2}\mathbf{A}_\Sigma^2\nabla(\Psi^*\Psi) = -\frac{\hbar^2}{2m_e}\frac{\partial}{\partial x_s}\left(\frac{\partial\Psi^*}{\partial x_s}\nabla\Psi + (\nabla\Psi^*)\frac{\partial\Psi}{\partial x_s}\right) + \frac{\hbar^2}{2m_e}\nabla\frac{\partial\Psi^*}{\partial x_s}\frac{\partial\Psi}{\partial x_s} + \frac{1}{c}\frac{\partial}{\partial x_s}\left(A_{\Sigma s}\mathbf{j}_0 + \right.$$

$$\left.\frac{e^2}{m_e c}A_{\Sigma s}\mathbf{A}_\Sigma\Psi^*\Psi\right) - \frac{1}{c}\nabla\left(j_{0s}A_{\Sigma s} + \frac{e^2}{m_e c}A_{\Sigma s}A_{\Sigma s}\Psi^*\Psi\right) + \frac{1}{c}\left(j_{0s} + \frac{e^2}{m_e c}A_{\Sigma s}\Psi^*\Psi\right)\nabla A_{\Sigma s} +$$

$$\frac{e^2}{2m_e c^2}\mathbf{A}_\Sigma^2\nabla(\Psi^*\Psi) = -\frac{\hbar^2}{2m_e}\frac{\partial}{\partial x_s}\left(\frac{\partial\Psi^*}{\partial x_s}\nabla\Psi + (\nabla\Psi^*)\frac{\partial\Psi}{\partial x_s}\right) + \frac{\hbar^2}{2m_e}\nabla\frac{\partial\Psi^*}{\partial x_s}\frac{\partial\Psi}{\partial x_s} + \frac{\partial}{\partial x_s}\left(\frac{1}{c}\mathbf{j}_0 A_{\Sigma s} + \right.$$

$$\left.\frac{e^2}{m_e c^2}\mathbf{A}_\Sigma A_{\Sigma s}\Psi^*\Psi\right) - \nabla\left(\frac{1}{c}\mathbf{j}_0\mathbf{A}_\Sigma + \frac{e^2}{2m_e c^2}\mathbf{A}_\Sigma^2\Psi^*\Psi\right) + \frac{1}{c}j_{0s}\nabla A_{\Sigma s}$$

Then (E2) takes the form

$$-i\hbar\left(\frac{\partial\Psi^*}{\partial t}\right)\nabla\Psi + i\hbar(\nabla\Psi^*)\frac{\partial\Psi}{\partial t} =$$

$$-\frac{\partial}{\partial x_k}\left(\frac{\hbar^2}{2m_e}\left(\frac{\partial\Psi^*}{\partial x_k}\nabla\Psi + (\nabla\Psi^*)\frac{\partial\Psi}{\partial x_k}\right) - \frac{1}{c}\mathbf{j}_0 A_{\Sigma k} - \frac{e^2}{m_e c^2}\mathbf{A}_\Sigma A_{\Sigma k}\Psi^*\Psi\right) -$$

$$\nabla\left(e\varphi_\Sigma\Psi^*\Psi - \mu_B H_{\Sigma s}\Psi^*\sigma_s\Psi + eG_s\Psi^*\sigma_s\Psi + \frac{1}{c}(\mathbf{j}_0\mathbf{A}_\Sigma) + \frac{e^2}{2m_e c^2}\mathbf{A}_\Sigma^2\Psi^*\Psi - \frac{\hbar^2}{2m_e}\frac{\partial\Psi^*}{\partial x_s}\frac{\partial\Psi}{\partial x_s}\right) +$$

$$e\Psi^*\Psi\nabla\varphi_\Sigma + \Psi^*\sigma_s\Psi\nabla(-\mu_B H_{\Sigma s} + eG_s) + \frac{1}{c}j_{0s}\nabla A_{\Sigma s} \quad (E3)$$

Similarly,

$$-i\hbar\frac{\partial\Psi^*}{\partial t}\Psi + i\hbar\frac{\partial\Psi}{\partial t}\Psi^* = \left(\frac{1}{2m_e}\left(-\frac{\hbar}{i}\nabla + \frac{e}{c}\mathbf{A}_\Sigma\right)^2\Psi^*\right)\Psi + \frac{1}{2m_e}\Psi^*\left(\frac{\hbar}{i}\nabla + \frac{e}{c}\mathbf{A}_\Sigma\right)^2\Psi - 2e\varphi_\Sigma\Psi^*\Psi -$$

$$2e\Psi^*\boldsymbol{\sigma}\Psi\mathbf{G} + \frac{e\hbar}{m_e c}\Psi^*\boldsymbol{\sigma}\Psi\mathbf{H}_\Sigma \quad (E4)$$

Let us transform, taking into account (40):

$$\left(\frac{1}{2m_e}\left(-\frac{\hbar}{i}\nabla + \frac{e}{c}\mathbf{A}_\Sigma\right)^2\Psi^*\right)\Psi + \frac{1}{2m_e}\Psi^*\left(\frac{\hbar}{i}\nabla + \frac{e}{c}\mathbf{A}_\Sigma\right)^2\Psi = -\frac{\hbar^2}{2m_e}(\Delta\Psi^*)\Psi - \frac{\hbar^2}{2m_e}\Psi^*\Delta\Psi -$$

$$\frac{e\hbar}{m_e c i}\left((\mathbf{A}_\Sigma\nabla)\Psi^*\right)\Psi + \frac{e\hbar}{m_e c i}\Psi^*(\mathbf{A}_\Sigma\nabla)\Psi + \frac{e^2}{m_e c^2}\mathbf{A}_\Sigma^2\Psi^*\Psi = -\frac{\hbar^2}{2m_e}\frac{\partial}{\partial x_s}\left(\frac{\partial\Psi^*}{\partial x_s}\Psi + \Psi^*\frac{\partial\Psi}{\partial x_s}\right) +$$

$$\frac{\hbar^2}{m_e}\frac{\partial\Psi^*}{\partial x_s}\frac{\partial\Psi}{\partial x_s} - \frac{e\hbar}{m_e c i}\mathbf{A}_\Sigma\left((\nabla\Psi^*)\Psi - \Psi^*\nabla\Psi\right) + \frac{e^2}{m_e c^2}\mathbf{A}_\Sigma^2\Psi^*\Psi = -\frac{\hbar^2}{2m_e}\frac{\partial}{\partial x_s}\left(\frac{\partial\Psi^*}{\partial x_s}\Psi + \Psi^*\frac{\partial\Psi}{\partial x_s}\right) +$$

$$\frac{\hbar^2}{m_e}\frac{\partial\Psi^*}{\partial x_s}\frac{\partial\Psi}{\partial x_s} - \frac{2}{c}(\mathbf{j}_0\mathbf{A}_\Sigma) - \frac{e^2}{m_e c^2}\mathbf{A}_\Sigma^2\Psi^*\Psi$$

Then (E4) takes the form

$$-i\hbar\frac{\partial\Psi^*}{\partial t}\Psi + i\hbar\frac{\partial\Psi}{\partial t} = -\frac{\hbar^2}{2m_e}\Delta(\Psi^*\Psi) + \frac{\hbar^2}{m_e}\frac{\partial\Psi^*}{\partial x_s}\frac{\partial\Psi}{\partial x_s} - \frac{2}{c}(\mathbf{j}_0\mathbf{A}_\Sigma) - \frac{e^2}{m_e c^2}\mathbf{A}_\Sigma^2\Psi^*\Psi - 2e\varphi_\Sigma\Psi^*\Psi -$$

$$2e\Psi^*\boldsymbol{\sigma}\Psi\mathbf{G} + \frac{e\hbar}{m_e c}\Psi^*\boldsymbol{\sigma}\Psi\mathbf{H}_\Sigma \tag{E5}$$

Using (E3), (E4) and (37), we transform (E1):

$$-\frac{m_e}{e}\frac{\partial\mathbf{j}_0}{\partial t} =$$

$$-\frac{\partial}{\partial x_k}\left(\frac{\hbar^2}{2m_e}\left(\frac{\partial\Psi^*}{\partial x_k}\nabla\Psi + (\nabla\Psi^*)\frac{\partial\Psi}{\partial x_k}\right) - \frac{1}{c}\mathbf{j}_0 A_{\Sigma k} - \frac{e^2}{m_e c^2}\mathbf{A}_\Sigma A_{\Sigma k}\Psi^*\Psi\right) - \nabla\left(-\frac{\hbar^2}{4m_e}\Delta(\Psi^*\Psi)\right) +$$

$$\rho\mathbf{E}_\Sigma + \frac{1}{c}j_{0s}\nabla A_{\Sigma s} + \frac{1}{c}\mathbf{A}_\Sigma\mathrm{div}\mathbf{j}_0 + (\Psi^*\sigma_s\Psi)\nabla(-\mu_B H_{\Sigma s} + eG_s) \tag{E6}$$

Taking into account the vector identity

$$j_{0s}\nabla A_{\Sigma s} = \mathbf{j}_0\times\mathrm{rot}\mathbf{A}_\Sigma + (\mathbf{j}_0\nabla)\mathbf{A}_\Sigma$$

one obtains

$$j_{0s}\nabla A_{\Sigma s} + \mathbf{A}_\Sigma\mathrm{div}\mathbf{j}_0 = \mathbf{j}_0\times\mathrm{rot}\mathbf{A}_\Sigma + (\mathbf{j}_0\nabla)\mathbf{A}_\Sigma + \mathbf{A}_\Sigma\mathrm{div}\mathbf{j}_0 = \mathbf{j}_0\times\mathbf{H}_\Sigma + \frac{\partial j_{0s}\mathbf{A}_\Sigma}{\partial x_s} \tag{E7}$$

Using (E7), we rewrite (E6) as

$$-\frac{m_e}{e}\frac{\partial j_{0i}}{\partial t} = -\frac{\partial \Pi_{ik}}{\partial x_k} + \rho E_{\Sigma i} + \frac{1}{c}(\mathbf{j}_0\times\mathbf{H}_\Sigma)_i + (\Psi^*\sigma_s\Psi)\frac{\partial}{\partial x_i}(-\mu_B H_{\Sigma s} + eG_s) \tag{E8}$$

where

$$\Pi_{ik} = \frac{\hbar^2}{2m_e}\left(\frac{\partial\Psi^*}{\partial x_k}\frac{\partial\Psi}{\partial x_i} + \frac{\partial\Psi^*}{\partial x_i}\frac{\partial\Psi}{\partial x_k}\right) - \frac{1}{c}(j_{0i}A_{\Sigma k} + j_{0k}A_{\Sigma i}) - \frac{e^2}{m_e c^2}A_{\Sigma i}A_{\Sigma k}\Psi^*\Psi - \delta_{ik}\frac{\hbar^2}{4m_e}\Delta(\Psi^*\Psi) \tag{E9}$$

Differentiating (112) with respect to time, one obtains

$$\frac{\partial\mathbf{P}_\Sigma}{\partial t} = \frac{\partial}{\partial t}\left(\frac{1}{4\pi c}\mathbf{E}_\Sigma\times\mathbf{H}_\Sigma\right) - \frac{m_e}{e}\frac{\partial\mathbf{j}_0}{\partial t} + \frac{\partial}{\partial t}\left(-\frac{1}{4\pi c^2}\frac{\partial G_k}{\partial t}\nabla G_k\right) + \frac{\hbar}{4}\mathrm{rot}\frac{\partial(\Psi^*\boldsymbol{\sigma}\Psi)}{\partial t} + \frac{1}{8\pi c^2}\mathrm{rot}\frac{\partial}{\partial t}\left(\mathbf{G}\times\frac{\partial\mathbf{G}}{\partial t}\right) \tag{E10}$$

Using (120), (122), (123), and (C1), one writes (E10) as

$$\frac{\partial P_{\Sigma i}}{\partial t} =$$

$$-\frac{\partial}{\partial x_k}\frac{1}{4\pi}\left(-E_{\Sigma i}E_{\Sigma k} - H_{\Sigma i}H_{\Sigma k} + \frac{1}{2}\delta_{ik}(\mathbf{E}_\Sigma^2 + \mathbf{H}_\Sigma^2) + \Pi_{ik}\right) - \frac{\partial}{\partial x_k}\frac{1}{4\pi}\left(\frac{\partial G_s}{\partial x_k}\frac{\partial G_s}{\partial x_i} - \delta_{ik}\left(\frac{1}{2}\frac{\partial G_s}{\partial x_m}\frac{\partial G_s}{\partial x_m} - \right.\right.$$

$$\left.\left.\frac{1}{2c^2}\frac{\partial G_s}{\partial t}\frac{\partial G_s}{\partial t}\right)\right) - \left((\rho_\Sigma - \rho)\mathbf{E}_\Sigma + \frac{1}{c}(\mathbf{j}_\Sigma - \mathbf{j}_0)\times\mathbf{H}_\Sigma\right)_i - \mu_B(\Psi^*\sigma_s\Psi)\frac{\partial H_{\Sigma s}}{\partial x_i} + \frac{\hbar}{4}\mathrm{rot}\frac{\partial(\Psi^*\boldsymbol{\sigma}\Psi)}{\partial t} +$$

$$\frac{1}{8\pi c^2}\mathrm{rot}\frac{\partial}{\partial t}\left(\mathbf{G}\times\frac{\partial\mathbf{G}}{\partial t}\right) \tag{E11}$$

We transform using the vector identity and relations (21) and (41):

$$-\mu_B(\Psi^*\sigma_s\Psi)\frac{\partial H_{\Sigma s}}{\partial x_i} = -\mu_B\frac{\partial(\Psi^*\sigma_s\Psi)H_{\Sigma s}}{\partial x_i} + \mu_B H_{\Sigma s}\frac{\partial}{\partial x_i}(\Psi^*\sigma_s\Psi) = -\mu_B\frac{\partial(\Psi^*\sigma_s\Psi)H_{\Sigma s}}{\partial x_i} + \mu_B\big(\mathbf{H}_\Sigma\times$$

$$\mathrm{rot}(\Psi^*\boldsymbol{\sigma}\Psi)\big)_i + \mu_B(\mathbf{H}_\Sigma\nabla)(\Psi^*\sigma_i\Psi) = -\mu_B\frac{\partial(\Psi^*\boldsymbol{\sigma}\Psi)\mathbf{H}_\Sigma}{\partial x_i} - \frac{1}{c}(\mathbf{H}_\Sigma\times c\mathrm{rot}\mathbf{m})_i + \mu_B\frac{\partial}{\partial x_k}(H_{\Sigma k}\Psi^*\sigma_i\Psi)$$

$$\tag{E12}$$

Taking into account (39), (41), (E12) and (119), one obtains

$$\frac{\partial P_{\Sigma i}}{\partial t} + \frac{\partial \sigma_{ik}^{(\Sigma)}}{\partial x_k} = -Ze\delta(\mathbf{r}-\mathbf{r}_n)\left(\mathbf{E}_\Sigma + \frac{1}{c}\mathbf{V}\times\mathbf{H}_\Sigma\right)_i \tag{E13}$$

where

$$\sigma_{ik}^{(\Sigma)} = \frac{1}{4\pi}\left(-E_{\Sigma i}E_{\Sigma k} - H_{\Sigma i}H_{\Sigma k} + \frac{1}{2}\delta_{ik}(\mathbf{E}_\Sigma^2 + \mathbf{H}_\Sigma^2)\right) + \Pi_{ik} + \frac{1}{4\pi}\frac{\partial G_s}{\partial x_k}\frac{\partial G_s}{\partial x_i} - \delta_{ik}\frac{1}{8\pi}\left(\frac{\partial G_s}{\partial x_m}\frac{\partial G_s}{\partial x_m} - \frac{1}{c^2}\frac{\partial G_s}{\partial t}\frac{\partial G_s}{\partial t}\right) - \varepsilon_{iks}\frac{\hbar}{4}\frac{\partial(\Psi^*\sigma_s\Psi)}{\partial t} + \frac{1}{8\pi c^2}\left(\frac{\partial}{\partial t}\left(G_k\frac{\partial G_i}{\partial t}\right) - \frac{\partial}{\partial t}\left(G_i\frac{\partial G_k}{\partial t}\right)\right) + \delta_{ik}\mu_B(\Psi^*\boldsymbol{\sigma}\Psi)\mathbf{H}_\Sigma - \mu_B(\Psi^*\sigma_i\Psi)H_{\Sigma k} \tag{E14}$$

## F. Movement of atoms and ions in an external electromagnetic field

We calculate

$$\frac{\partial \mathbf{P}_a}{\partial t} = -\frac{m_e}{e}\frac{\partial \mathbf{j}_0}{\partial t} + \frac{\hbar}{4}\text{rot}\frac{\partial(\Psi^*\boldsymbol{\sigma}\Psi)}{\partial t} + m_n\dot{\mathbf{V}}\delta(\mathbf{r}-\mathbf{r}_n) - m_n\mathbf{V}(\mathbf{V}\nabla)\delta(\mathbf{r}-\mathbf{r}_n)$$

Using (123) and (127), one obtains

$$\frac{\partial P_{ai}}{\partial t} = -\frac{\partial}{\partial x_k}\left(\Pi_{ik} + m_n V_i V_k \delta(\mathbf{r}-\mathbf{r}_n) - \varepsilon_{iks}\frac{\hbar}{4}\frac{\partial(\Psi^*\sigma_s\Psi)}{\partial t}\right) + \rho E_{\Sigma i} + \frac{1}{c}(\mathbf{j}_0\times\mathbf{H}_\Sigma)_i$$
$$+ (\Psi^*\sigma_s\Psi)\frac{\partial}{\partial x_i}(-\mu_B H_{\Sigma s} + eG_s) + Ze\delta(\mathbf{r}-\mathbf{r}_n)\left(\mathbf{E}_\Sigma + \frac{1}{c}\mathbf{v}_n\times\mathbf{H}_\Sigma\right)_i$$

Using (E12) and (39), one obtains

$$\frac{\partial P_{ai}}{\partial t} + \frac{\partial K_{ik}}{\partial x_k} = \left(\rho_\Sigma \mathbf{E}_\Sigma + \frac{1}{c}\mathbf{j}_\Sigma\times\mathbf{H}_\Sigma\right)_i + e(\Psi^*\sigma_s\Psi)\frac{\partial G_s}{\partial x_i} \tag{F1}$$

where

$$K_{ik} = \Pi_{ik} + m_n V_i V_k \delta(\mathbf{r}-\mathbf{r}_n) - \varepsilon_{iks}\frac{\hbar}{4}\frac{\partial(\Psi^*\sigma_s\Psi)}{\partial t} + \delta_{ik}\mu_B(\Psi^*\boldsymbol{\sigma}\Psi)\mathbf{H}_\Sigma - \mu_B H_{\Sigma k}\Psi^*\sigma_i\Psi \tag{F2}$$

Consider

$$(\mathbf{j}_\Sigma\times(\boldsymbol{\xi}\nabla)\mathbf{H})_i = \left(\mathbf{j}_\Sigma\times\frac{\partial\mathbf{H}}{\partial x_s}\xi_s\right)_i = \varepsilon_{ikm}j_{\Sigma k}\xi_s\frac{\partial H_m}{\partial x_s}$$

where $\boldsymbol{\xi} = \mathbf{r} - \mathbf{r}_n$.

$$j_{\Sigma k}\xi_s = \frac{1}{2}(j_{\Sigma k}\xi_s - j_{\Sigma s}\xi_k) + \frac{1}{2}(j_{\Sigma k}\xi_s + j_{\Sigma s}\xi_k)$$

$$j_{\Sigma k}\xi_s + j_{\Sigma s}\xi_k = (\mathbf{j}_\Sigma\nabla)(\xi_s\xi_k) = \nabla(\mathbf{j}_\Sigma\xi_s\xi_k) - \xi_s\xi_k\nabla\mathbf{j}_\Sigma = \nabla(\mathbf{j}_\Sigma\xi_s\xi_k) + \xi_s\xi_k\frac{\partial\rho_\Sigma}{\partial t}$$

$$(\boldsymbol{\xi}\times\mathbf{j}_\Sigma)_i = \varepsilon_{ilm}\xi_l j_{\Sigma m}$$

$$\varepsilon_{rsk}(\boldsymbol{\xi}\times\mathbf{j}_\Sigma)_r = \varepsilon_{rsk}\varepsilon_{rlm}\xi_l j_{\Sigma m} = j_{\Sigma k}\xi_s - j_{\Sigma s}\xi_k$$

$$j_{\Sigma k}\xi_s - j_{\Sigma s}\xi_k = \varepsilon_{rsk}(\boldsymbol{\xi}\times\mathbf{j}_\Sigma)_r$$

Thus

$$j_{\Sigma k}\xi_s = \frac{1}{2}\varepsilon_{rsk}(\boldsymbol{\xi}\times\mathbf{j}_\Sigma)_r + \frac{1}{2}\nabla(\mathbf{j}_\Sigma\xi_s\xi_k) + \frac{1}{2}\xi_s\xi_k\frac{\partial\rho_\Sigma}{\partial t}$$

Taking into account that $\text{div}\mathbf{H} = 0$, one obtains

$$(\mathbf{j}_\Sigma \times (\boldsymbol{\xi}\nabla)\mathbf{H})_i = \frac{1}{2}(\boldsymbol{\xi} \times \mathbf{j}_\Sigma)_m \frac{\partial H_m}{\partial x_i} + \varepsilon_{ikm}\frac{1}{2}\nabla(\mathbf{j}_\Sigma \xi_s \xi_k)\frac{\partial H_m}{\partial x_s} + \frac{1}{2}\varepsilon_{ikm}\xi_s\xi_k \frac{\partial H_m}{\partial x_s}\frac{\partial \rho_\Sigma}{\partial t}$$

Then, taking into account that in the absence of ionization $\oint \xi_s \xi_k \mathbf{j}_\Sigma d\mathbf{f} = 0$, one obtains

$$\int (\mathbf{j}_\Sigma \times ((\mathbf{r}-\mathbf{r}_n)\nabla)\mathbf{H})_i dV = \frac{\partial \mathbf{H}}{\partial x_i}\int \frac{1}{2}(\mathbf{r}-\mathbf{r}_n) \times \mathbf{j}_\Sigma dV + \frac{1}{2}\varepsilon_{ikm}\frac{\partial H_m}{\partial x_s}\int \xi_s\xi_k \frac{\partial \rho_\Sigma}{\partial t} dV \quad (F3)$$

Taking into account (140), one obtains

$$\int \xi_s\xi_k \frac{\partial \rho_\Sigma}{\partial t} dV = \frac{d}{dt}\int \xi_s\xi_k \rho_\Sigma dV - \int \rho_\Sigma \frac{\partial \xi_s\xi_k}{\partial t} dV =$$

$$\frac{d}{dt}\int \xi_s\xi_k \rho_\Sigma dV + V_s \int \rho_\Sigma \xi_k dV + V_k \int \rho_\Sigma \xi_s dV = \frac{1}{3}\dot{D}_{ks} + \delta_{sk}\frac{1}{3}\frac{d}{dt}\int \boldsymbol{\xi}^2 \rho_\Sigma dV + V_s d_{\Sigma k} + V_k d_{\Sigma s} \quad (F4)$$

where

$$D_{ks} = \int (3\xi_s\xi_k - \boldsymbol{\xi}^2 \delta_{sk})\rho_\Sigma dV \quad (F5)$$

is the quadrupole moment of the electron field with respect to the nucleus.

Taking into account (38) and (F4), one writes (F3) as

$$\int (\mathbf{j}_\Sigma \times ((\mathbf{r}-\mathbf{r}_n)\nabla)\mathbf{H})_i dV = c\mathbf{M}\frac{\partial \mathbf{H}}{\partial x_i} + \frac{1}{6}\varepsilon_{ikm}\frac{\partial H_m}{\partial x_s}\dot{D}_{ks} + \frac{1}{6}(\text{rot}\mathbf{H})_i \frac{d}{dt}\int \boldsymbol{\xi}^2 \rho_\Sigma dV + \frac{1}{2}(\mathbf{V}\nabla)(\mathbf{d}_\Sigma \times$$

$$\mathbf{H})_i + \frac{1}{2}(\mathbf{d}_\Sigma \nabla)(\mathbf{V} \times \mathbf{H})_i \quad (F6)$$

where the magnetic moment **M** is calculated with respect to the nucleus and, according to (44), is the sum of the convective (orbital) magnetic moment of the electron field $\mathbf{M}_{or}$, calculated with respect to the nucleus, and the intrinsic (spin) magnetic moment of the electron field **μ**.

If inside the atom (ion), there are no sources of external field **H**, the term rot**H** in (F6) can be replaced by $\frac{1}{c}\frac{\partial \mathbf{E}}{\partial t}$ according to Maxwell equation.

Then, taking into account (139)-(141) and (F6), one writes equation (138) as

$$\frac{\partial}{\partial t}\int \mathbf{P}_a dV = q_\Sigma \mathbf{E}_0 + q_\Sigma \frac{1}{c}\mathbf{V} \times \mathbf{H}_0 + (\mathbf{d}_\Sigma \nabla)\mathbf{E} + \nabla(\mathbf{MH}) + \int \left(\rho_\Sigma \mathbf{E}_a + \frac{1}{c}\mathbf{j}_\Sigma \times \mathbf{H}_a\right) dV +$$

$$\int e(\Psi^* \sigma_s \Psi)\nabla G_s dV + \mathbf{F}_{add} \quad (F7)$$

where

$$\mathbf{F}_{add} = \frac{1}{c}\dot{\mathbf{d}}_\Sigma \times \mathbf{H}_0 + \mathbf{F}_D + \frac{1}{6c^2}\frac{\partial \mathbf{E}}{\partial t}\frac{d}{dt}\int \boldsymbol{\xi}^2 \rho_\Sigma dV + \frac{1}{2c}(\mathbf{V}\nabla)(\mathbf{d}_\Sigma \times \mathbf{H}) + \frac{1}{2c}(\mathbf{d}_\Sigma \nabla)(\mathbf{V} \times \mathbf{H}) \quad (F8)$$

$$F_{Di} = \frac{1}{6c}\varepsilon_{ikm}\frac{\partial H_m}{\partial x_s}\dot{D}_{ks} \quad (F9)$$

The forces $\int \left(\rho_\Sigma \mathbf{E}_a + \frac{1}{c}\mathbf{j}_\Sigma \times \mathbf{H}_a\right) dV$ and $\int e(\Psi^* \sigma_s \Psi)\nabla G_{as} dV$, which describe the interaction of charges and currents inside an atom (ion) with its own electromagnetic field and its own **G**-field, created by an electron wave and a nucleus, are equal to zero, because they are the internal forces of the system.